\DeclareUnicodeCharacter{22C6}{\ensuremath{\star}} 
\DeclareUnicodeCharacter{2212}{\textminus} 
\documentclass[a4paper,11pt]{article}
\pdfoutput=1 
\usepackage{jcappub} 
\usepackage[T1]{fontenc} 

\usepackage{float, graphicx, amsmath, amssymb, amsthm, bm, bbm, xfrac}

\usepackage{hyperref}        
\usepackage{listings, color} 
\usepackage{esint}           
\usepackage{mathtools}       
\usepackage{tensor}          
\usepackage{xspace}          
\usepackage[scr=esstix]{mathalfa} 
\usepackage{stackengine} 
\usepackage{orcidlink} 
\usepackage{makecell,booktabs} 

\usepackage{makecell}

\usepackage{textgreek}

\setcellgapes{2pt}
\makegapedcells

\graphicspath{{./figures/}}

\newcommand{\Fisher}[0]{\mathcal{I}}
\newcommand{\Fishermat}[1]{\left[\Fisher\left({#1}\right)\right]}
\newcommand{\Fisherel}[3]{\Fishermat{{#1}}_{{#2}{#3}}}
\newcommand{\Fisherij}[1]{\Fisherel{{#1}}{i}{j}}

\usepackage{xstring}

\newcommand{\pp}[3]{
	\IfStrEq{#2}{#3}{
		\frac{\partial^2{#1}}{\partial{#2}^2}
	}{
		\frac{\partial^2{#1}}{\partial{#2}\partial{#3}}
	}
}
\newcommand{\ppp}[4]{
	\IfStrEq{#2}{#3}{
		\IfStrEq{#3}{#4}{
			\frac{\partial^3{#1}}{\partial{#2}^3}
		}{
			\frac{\partial^3{#1}}{\partial{#2}^2\partial{#4}}
		}
	}{
		\IfStrEq{#3}{#4}{
			\frac{\partial^3{#1}}{\partial{#2}\partial{#3}^2}
		}{
			\frac{\partial^3{#1}}{\partial{#2}\partial{#3}\partial{#4}}
		}
	}
}

\newcommand{\simin}{ \mathrel{\underset{\sim}{\in}} }

\newtoggle{usedEFT}\togglefalse{usedEFT}
\newtoggle{usedGRF}\togglefalse{usedGRF}
\newtoggle{usedHEC}\togglefalse{usedHEC}
\newtoggle{usedKLD}\togglefalse{usedKLD}
\newtoggle{usedLPRE}\togglefalse{usedLPRE}
\newtoggle{usedMCMC}\togglefalse{usedMCMC}
\newtoggle{usedNL}\togglefalse{usedNL}
\newtoggle{usedPDF}\togglefalse{usedPDF}
\newtoggle{usedPRE}\togglefalse{usedPRE}
\newtoggle{usedVID}\togglefalse{usedVID}

\newtoggle{usedBAO}\togglefalse{usedBAO}
\newtoggle{usedCIB}\togglefalse{usedCIB}
\newtoggle{usedCMB}\togglefalse{usedCMB}
\newtoggle{usedDM}\togglefalse{usedDM}
\newtoggle{usedELG}\togglefalse{usedELG}
\newtoggle{usedEoR}\togglefalse{usedEoR}
\newtoggle{usedHI}\togglefalse{usedHI}
\newtoggle{usedHMF}\togglefalse{usedHMF}
\newtoggle{usedISM}\togglefalse{usedISM}
\newtoggle{usedISW}\togglefalse{usedISW}
\newtoggle{usedLIM}\togglefalse{usedLIM}
\newtoggle{usedLRG}\togglefalse{usedLRG}
\newtoggle{usedLSS}\togglefalse{usedLSS}
\newtoggle{usedSZ}\togglefalse{usedSZ}
\newtoggle{usedtSZ}\togglefalse{usedtSZ}
\newtoggle{usedkSZ}\togglefalse{usedkSZ}
\newtoggle{usedWGL}\togglefalse{usedWGL}

\newtoggle{usedNG}\togglefalse{usedNG}
\newtoggle{usedPING}\togglefalse{usedPING}
\newtoggle{usedPNG}\togglefalse{usedPNG}

\newtoggle{usedCCAT}\togglefalse{usedCCAT}
\newtoggle{usedCOMAP}\togglefalse{usedCOMAP}
\newtoggle{usedDSS}\togglefalse{usedDSS}
\newtoggle{usedEoRSpec}\togglefalse{usedEoRSpec}
\newtoggle{usedFYST}\togglefalse{usedFYST}
\newtoggle{usedStwo}\togglefalse{usedStwo}

\newcommand{\EFT}{%
    \iftoggle{usedEFT}%
        {EFT\xspace}%
        {\toggletrue{usedEFT}effective field theory (EFT)\xspace}%
}
\newcommand{\GRF}{%
    \iftoggle{usedGRF}%
        {GRF\xspace}%
        {\toggletrue{usedGRF}Gaussian random field (GRF)\xspace}%
}
\newcommand{\HEC}{%
    \iftoggle{usedHEC}%
        {HEC\xspace}%
        {\toggletrue{usedHEC}homogeneous ellipsoidal collapse (HEC)\xspace}%
}
\newcommand{\KLD}{%
    \iftoggle{usedKLD}%
        {KLD\xspace}%
        {\toggletrue{usedKLD}Kullback-Leibler divergence (KLD)\xspace}%
}
\newcommand{\LPRE}{%
    \iftoggle{usedLPRE}%
        {LPRE\xspace}%
        {\toggletrue{usedLPRE}logarithmic point-wise relative entropy (LPRE)\xspace}%
}
\newcommand{\MCMC}{%
    \iftoggle{usedMCMC}%
        {MCMC\xspace}%
        {\toggletrue{usedMCMC}Markov chain Monte Carlo (MCMC)\xspace}%
}
\newcommand{\NL}{%
    \iftoggle{usedNL}%
        {NL\xspace}%
        {\toggletrue{usedNL}non-linear (NL)\xspace}%
}
\newcommand{\PDF}{%
    \iftoggle{usedPDF}
        {PDF\xspace}%
        {\toggletrue{usedPDF}probability density function (PDF)\xspace}%
}
\newcommand{\PRE}{%
    \iftoggle{usedPRE}%
        {PRE\xspace}%
        {\toggletrue{usedPRE}point-wise relative entropy (PRE)\xspace}%
}
\newcommand{\VID}{%
    \iftoggle{usedVID}%
        {VID\xspace}%
        {\toggletrue{usedVID}voxel intensity distribution (VID)\xspace}%
}

\newcommand{\BAO}{%
    \iftoggle{usedBAO}%
        {BAO\xspace}%
        {\toggletrue{usedBAO}baryon acoustic oscillations (BAO)\xspace}%
}
\newcommand{\CIB}{%
    \iftoggle{usedCIB}%
        {CIB\xspace}%
        {\toggletrue{usedCIB}cosmic infrared background (CIB)\xspace}%
}
\newcommand{\CMB}{%
    \iftoggle{usedCMB}%
        {CMB\xspace}%
        {\toggletrue{usedCMB}cosmic microwave background (CMB)\xspace}%
}
\newcommand{\DM}{%
    \iftoggle{usedDM}%
        {DM\xspace}%
        {\toggletrue{usedDM}dark matter (matter that interacts at most very weakly with light; DM)\xspace}%
}

\newcommand{\EoR}{%
    \iftoggle{usedEoR}%
        {EoR\xspace}%
        {\toggletrue{usedEoR}epoch of reionisation (EoR)\xspace}%
}

\newcommand{\HMF}{%
    \iftoggle{usedHMF}%
        {HMF\xspace}%
        {\toggletrue{usedHMF}halo mass function (HMF)\xspace}%
}
\newcommand{\ISM}{%
    \iftoggle{usedISM}%
        {ISM\xspace}%
        {\toggletrue{usedISM}interstellar medium (ISM)\xspace}%
}
\newcommand{\ISW}{%
    \iftoggle{usedISW}%
        {ISW\xspace}%
        {\toggletrue{usedISW}integrated Sachs\textendash{}Wolfe (ISW)\xspace}%
}
\newcommand{\LambdaCDM}{$\Lambda$CDM\xspace}
\NewDocumentCommand{\LIM}{s}{%
    \iftoggle{usedLIM}{%
        LIM\xspace%
    }{%
        \toggletrue{usedLIM}%
        \IfBooleanTF{#1}{%
            Line intensity mapping (LIM)\xspace%
        }{%
            line intensity mapping (LIM)\xspace%
        }%
    }%
}
\newcommand{\LRG}{%
    \iftoggle{usedLRG}%
        {LRG\xspace}%
        {\toggletrue{usedLRG}luminous red galaxy (LRG)\xspace}%
}
\newcommand{\LRGs}{%
    \iftoggle{usedLRG}%
        {LRGs\xspace}%
        {\toggletrue{usedLRG}luminous red galaxies (LRGs)\xspace}%
}
\newcommand{\LSS}{%
    \iftoggle{usedLSS}%
        {LSS\xspace}%
        {\toggletrue{usedLSS}large-scale structure (LSS)\xspace}%
}
\newcommand{\SZ}{%
    \iftoggle{usedSZ}%
        {SZ\xspace}%
        {\toggletrue{usedSZ}Sunyaev\textendash{}Zel'dovich (SZ)\xspace}%
}
\newcommand{\tSZ}{%
    \iftoggle{usedtSZ}%
        {tSZ\xspace}%
        {\toggletrue{usedtSZ}thermal Sunyaev\textendash{}Zel'dovich (tSZ)\xspace}%
}
\newcommand{\tSZeffect}{%
    \iftoggle{usedtSZ}%
        {tSZ effect\xspace}%
        {\toggletrue{usedtSZ}thermal Sunyaev\textendash{}Zel'dovich effect (tSZ)\xspace}%
}

\NewDocumentCommand{\WGL}{s}{%
    \iftoggle{usedWGL}{%
        WGL\xspace%
    }{%
        \toggletrue{usedWGL}%
        \IfBooleanTF{#1}{%
            Weak gravitational lensing (WGL)\xspace%
        }{%
            weak gravitational lensing (WGL)\xspace
        }%
    }%
}

\newcommand{\nong}{%
    \iftoggle{usedNG}%
        {NG\xspace}%
        {\toggletrue{usedNG}non-Gaussianity (NG)\xspace}%
}
\newcommand{\ping}{%
    \iftoggle{usedPING}%
        {PING\xspace}%
        {\toggletrue{usedPING}primordial intermittent non-Gaussianity (PING)\xspace}%
}
\newcommand{\png}{%
    \iftoggle{usedPNG}%
        {PNG\xspace}%
        {\toggletrue{usedPNG}primordial non-Gaussianity (PNG)\xspace}%
}

\newcommand{\fNL}{f_\mathrm{NL}\xspace}
\newcommand{\fNLlocal}{f_\mathrm{NL}^\mathrm{local}\xspace}

\newcommand{\zetag}{\zeta_g}
\newcommand{\dzeta}{\Delta\zeta}
\newcommand{\dtildezeta}{\Delta\tilde{\zeta}}
\newcommand{\kpulse}{k_\mathrm{pulse}}

\newcommand{\cii}{$[\mathrm{C}\textsc{ii}]$\xspace}%
\newcommand{\hi}{$\mathrm{H}\textsc{i}$\xspace}%
\newcommand{\halpha}{$\mathrm{H}\alpha$\xspace}%
\newcommand{\oii}{$[\mathrm{O}\textsc{ii}]$\xspace}%
\newcommand{\oiii}{$[\mathrm{O}\textsc{iii}]$\xspace}%
\DeclareMathOperator{\alphacii}{\alpha_{[\mathrm{C\textsc{ii}}]}}
\DeclareMathOperator{\Lcii}{L_{[\mathrm{C\textsc{ii}}]}}
\DeclareMathOperator{\Icii}{I_{[\mathrm{C\textsc{ii}}]}}
\DeclareMathOperator{\Lciihalo}{L_{[\mathrm{C\textsc{ii}}],\mathrm{halo}}}
\DeclareMathOperator{\nucii}{\nu_{[\mathrm{C\textsc{ii}}],0}}
\DeclareMathOperator{\lambdacii}{\lambda_{[\mathrm{C\textsc{ii}}],0}}
\DeclareMathOperator{\mhi}{M_{\mathrm{H\textsc{i}}}}

\newcommand{\pkp}{\texttt{Peak Patch}\xspace}%
\newcommand{\ws}{\texttt{WebSky}\xspace}%
\newcommand{\wsone}{\texttt{WebSky1.0}\xspace}%
\newcommand{\cobaya}{\texttt{Cobaya}\xspace}%
\newcommand{\class}{\texttt{CLASS}\xspace}%
\newcommand{\nbody}{$N$-body\xspace}%
\newcommand{\hydro}{hydrodynamical\xspace}%
\newcommand{\CCAT}{%
    \iftoggle{usedCCAT}%
        {CCAT\xspace}%
        {\toggletrue{usedCCAT}Cerro Chajnantor Atacama Telescope (CCAT)\xspace}%
}

\newcommand{\DSS}{%
    \iftoggle{usedDSS}%
        {DSS\xspace}%
        {\toggletrue{usedDSS}Deep Spectroscopic Survey (DSS)\xspace}%
}
\newcommand{\planck}{\textit{Planck}\xspace}%
\newcommand{\EoRSpec}{%
    \iftoggle{usedEoRSpec}%
    {EoR-Spec\xspace}%
    {\togglefalse{usedEoRSpec}Epoch of Reionisation Spectrometer (EoR-Spec)\xspace}%
}
\newcommand{\FYST}{%
    \iftoggle{usedFYST}%
        {FYST\xspace}%
        {\toggletrue{usedFYST}Fred Young Submillimeter Telescope (FYST)\xspace}%
}
\newcommand{\Stwo}{%
    \iftoggle{usedStwo}%
        {S2\xspace}%
        {\toggletrue{usedStwo}Stage 2 LIM survey (S2)\xspace}%
}

\newcommand{\nl}{non-linear\xspace}

\defcitealias{horlaville2023}{HCBL}
\newtoggle{usedHCBL}\togglefalse{usedHCBL}
\newcommand{\citeHCBL}{%
    \iftoggle{usedHCBL}%
        {\citetalias{horlaville2023}\xspace}%
        {\toggletrue{usedHCBL}\cite{horlaville2023} (\citetalias{horlaville2023})\xspace}%
}

\defcitealias{morrison2024a}{MBB}
\newtoggle{usedMBB}\togglefalse{usedMBB}
\newcommand{\citeMBB}{%
    \iftoggle{usedMBB}%
        {\citetalias{morrison2024a}\xspace}%
        {\toggletrue{usedMBB}\cite{morrison2024a} (\citetalias{morrison2024a})\xspace}%
}

\newcommand \countme{ \addtocounter{equation}{1} \tag{\theequation} }

\title{The \texorpdfstring{\ws}{WebSky} \texorpdfstring{\textnormal{\cii}}{[CII]} Forecasts and the search for primordial intermittent non-Gaussianity}

\author[a,b,1]{Nathan J. Carlson\orcidlink{0000-0002-2731-7708}}
\author[a]{J. Richard Bond\orcidlink{0000-0003-2358-9949}}
\author[c]{Dongwoo T. Chung\orcidlink{0000-0003-2618-6504}}
\author[d,e]{Patrick Horlaville\orcidlink{0009-0007-3541-435X}}
\author[a,b]{Thomas Morrison}

\affiliation[a]{Canadian Institute for Theoretical Astrophysics, University of Toronto, 60 St. George Street, Toronto, ON, M5S 3H8, Canada}
\affiliation[b]{Department of Physics, University of Toronto, 60 St. George Street, Toronto, ON, M5S 3H8, Canada}
\affiliation[c]{Department of Astronomy, Cornell University, 122 Sciences Dr, Ithaca, NY 14850, United States}
\affiliation[d]{Department of Physics, McGill University, 3600 Rue University, Montreal, QC H3A 2T8, Canada}
\affiliation[e]{Trottier Space Institute, McGill University, 3550 Rue University, Montreal, QC, H3A 2A7, Canada}

\emailAdd{njcarlson@cita.utoronto.ca}

\abstract{We present the \ws \cii line-intensity mock maps and forecast the capabilities of upcoming wide-field submillimeter-wave surveys of cosmological \cii emission from the epoch of reionization (EoR). Using the \pkp algorithm to generate light-cone dark matter (DM) halo catalogues and the \ws framework to forward-model the cosmological \cii signal, we construct tomographic mock surveys matched to the CCAT Observatory. We investigate both astrophysical models of \cii emission from interstellar gas and the potential for the study of primordial intermittent non-Gaussianity (PING) as a science case for Stage 2 line intensity mapping (LIM) surveys. The \cii voxel intensity distribution (VID) is used as a summary statistic in forecasts. Additional constraints on PING are derived from a relative entropy study of \pkp halo mass functions. We show that upcoming LIM surveys will provide insights into the way we model cosmological line emission, and next-generation surveys can place competitive bounds on novel inflationary scenarios such as PING. The \ws \cii mocks and corresponding \pkp halo catalogues are publicly available at \href{https://uoft.me/webskycii}{https://uoft.me/webskycii}.}

\begin{document}
\maketitle
\raggedbottom

\section{Introduction}\label{sec:intro}

The largest distinct physical structures in the observable universe are dense clusters of gravitationally bound galaxies and the more diffuse filaments of galaxies that weave them together in the interconnected network known as the cosmic web \cite{Bond_1996}. The web-like network appeared early in the formation of cosmic structure, though its makeup has changed as the universe evolved. Over time, gravity funnelled galaxies along filaments into increasingly massive clusters, and the expansion of the universe drove galaxies in sparsely populated regions apart, leaving increasingly empty cosmic voids. The evolution of this \LSS is rich with information, understanding it throughout cosmic time has revealed, and will continue to reveal much about the fundamental laws of nature.

Observations and simulations of the evolution of the cosmic web are both qualitatively and quantitatively compatible with the standard model of cosmology, the so-called \LambdaCDM model \cite{ostriker1995,Riess_1998}, with a few tensions possibly indicating something beyond. In this standard model, the current evolution of the universe is dominated by dark energy that drives the overall large-scale expansion of the universe. Dark energy possibly takes the form of a uniform potential energy density, which can be represented by the cosmological constant $\Lambda$ in the field equations for General Relativity. Meanwhile, clustering in \LambdaCDM, whether in the present epoch of dark energy dominance or not, is driven by cold, non-relativistic \DM. Along with the non-relativistic baryonic matter, the \DM determines the spatially varying and deepening gravitational potential wells which seed and amplify the gravitational collapse of the overdense regions that make up the back bone of the cosmic web \citep{BST_1982,Peebles_1982,blumenthal_1984}.

At earlier times, \LambdaCDM posits that the universe was hot and ionized. As it expanded and cooled, it underwent a phase transition known as recombination, forming neutral atoms \cite{Zeldovich_1968,Peebles_1968}, allowing photons to decouple from the baryonic matter and nearly free-stream to us as the \CMB \citep{Alpher_Bethe_Gamow_1948,Alpher_1948,Alpher_1949,Dicke_1965}. Modern measurements of the \CMB \citep{planckcollaboration_2018_I,ACT_2025} show that it is distributed as a nearly-Gaussian random field, and thus, before the formation of the cosmic web, the baryons and \DM were also well-modelled by \GRF statistics \citep{bbks}. Once expansive Hubble drag ceased to be dominated by photons and neutrinos \DM started to cluster, so by the time the \CMB was released, this clustering had entered an extended linear phase. The familiar cosmic web patterns appeared as the fluctuations approached mild non-linearity, and eventually dense \nong structures such as galaxies and stars began to form \cite{Zeldovich_1970,Bond_1996}.

The recent evolution of cosmic structure can be pieced together from galaxy and quasar surveys that map our cosmic neighbourhood \citep{SDSS-IV_2021,DESI_2025,bechtol_2025,DES_Collaboration_2025}. These surveys support the hypothesis that galaxies grow over time by accreting matter and merging, thus younger galaxies at higher redshift tend to be of lower mass, containing fewer stars, and have lower luminosity. In an expanding universe, surface luminosity is proportional to $(1+z)^{-4}$ \cite{Tolman_1930}, so galaxies at high redshift are much dimmer, requiring more sensitivity and observing time to resolve. Matters are complicated further as light from galaxies at redshift greater than a few are Doppler shifted into bands that are contaminated by atmospheric sources \cite{Rousselot_2000,Ellis_2020}, hindering dedicated ground-based surveys. As a result, galaxy catalogues are much less complete above redshifts of a few. This motivates surveys of bulk properties of the cosmic web rather than individual galaxies to piece together the billion or so years of evolution between the \CMB and the regimes in which our galaxy catalogues are complete.
 
Disturbance of the free-streaming \CMB photons on their way to us allows for the inference of some structure, but these are projected effects that are integrated along lines of sight with most of their contributions sourced from relatively nearby effects ($z\lesssim5$). During the \EoR ($5 \lesssim z \lesssim 15$), the first stars are theorized to form and re-ionize their surroundings, which would then glow in characteristic emission line radiation. Emission lines are highly localized in frequency, so redshifts of cosmological line emission sources can be precisely determined spectroscopically. \LIM* is a technique that uses spectrographs to produce three dimensional maps of line emission structures as they evolve with redshift. While isolating line emission from foreground contamination is a challenge, this emerging field is the subject of many pathfinder experiments. These include surveys of recombination lines \halpha, \oii and \oiii \citep{silva_2017,dore_2015,hill_2008}, 21-cm \hi \citep{chime_collaboration_2022,wang_2021,newburgh_2016,vanderlinde_2020}, millimetre-wave CO \citep{Cleary_2022,karkare_2022b} or, the focus of this work, submillimetre \cii \citep{CCAT_Prime_Collaboration_2022,sun_2021,Essinger_Hileman_2020,fasano_2022,vieira_2020}. A more complete understanding of the stellar environments emitting the cosmological signal is required to distinguish it from these foregrounds, necessitating further study. 

The origins of the primordial cosmological fields and the correlations observed in the \CMB may be the result of an inflationary phase in the very early universe. In this scenario, the potential energy density swamps the kinetic energy density, leading to dynamics dominated by a negative pressure, which results in accelerated expansion \cite{Starobinsky_1980,guth_1981,Linde_1982}. There are many effective potential energy densities that can give rise to such acceleration, and a top-down theory of inflation from first principles remains elusive. Inflation is a phenomenology rooted in the behaviour in the Hubble parameter, $H$, the acceleration (or equation of state) parameter, $\epsilon = -d\ln H/d\ln a$, and its fluctuations \cite{Albrecht_1982,Salopek_1990,mukhanov_1992}. The landscape for viable inflation models is wide, and invariably involves fields beyond the so-called inflaton, $\phi$, which can be thought of as a local Hubble parameter, $H(\mathbf{x},t)$, or the related total energy density. The total effective potential, $3M_P^2 H^2 (1-\epsilon /3)$, total kinetic energy, $3M_P^2 H^2 \epsilon /3$, and the inflation/deceleration boundary, $\epsilon=1$, allow a richness of possibilities in additional fields, which leads to a diversity in inflation models. 


One obvious thing to look for is a breakdown of the nearly Gaussian initial conditions of the \LambdaCDM model. Usually such deviations have been modelled by a perturbative approach to \png characterized by a single parameter, $\fNL$, multiplying a specific template for 3-point fluctuations that are not present in the primordially Gaussian case \cite{Maldacena_2003,Dalal_2008}. \CMB measurements have placed relatively strong constraints on $\fNL$ \cite{planckcollaboration_2018_IX}, but still far off what might be expected from most inflaton models. More importantly, other forms of the \png that are quite generic may arise, characterized by localization in position space and in wavenumber space, which we describe as ``intermittency''. A related work, \citeMBB, describes \ping in which fields transverse to the mean flow of the effective potential lead to the appearance of radical modifications to the nearly-Gaussian initial conditions assumed in \LambdaCDM, much beyond what conventional historical inflation models produce \cite{Starobinsky_1980,guth_1981,Lucchin_1984,Freese_1990,
Boubekeur_2005}. The temptation is to extrapolate in the simplest possible way from the nine $e$-folds in wavenumber that \CMB and \LSS observations constrain on \png to the 50 $e$-folds below this range that covers the much less constrained early galaxy formation epochs. If perturbative \nong of the $\fNL\lesssim5$ template form are used with the current constraints from \CMB \cite{planckcollaboration_2018_IX}, little difference is found in galaxy formation from the pure Gaussian $\fNL=0$ case, hence we do not concentrate our \LSS analysis of the \LIM signal on that scenario. Instead, our focus is characterizing non-perturbative \ping, and trying to find evidence for it buried in the developing statistics of early galaxies, as we strive for a complete theory of \emph{all scales} in the inflationary phase. 

Gravitational dynamics, \LIM effects and other \NL astrophysical processes produce \nong with distinct forms that must be distinguished from \png generated during the inflationary epoch. To disentangle them, we employ mock sky maps, simulated pictures of the sky analogous to observations with a specific instrument. Mocks attempt to accurately portray foregrounds and instrumental noise so that deviations from a fiducial cosmological model (such as novel \png) can be compared against observations. In this work, we focus on \CCAT \citep{CCAT_Prime_Collaboration_2022} surveys of \cii, and its future extensions, but our methodology is transferable to other observatories and other lines as well.

The cosmological \LIM signal is emitted in the \ISM as interstellar gas is excited by stellar radiation and emits in characteristic energies. The \LIM signal is therefore a tracer of the mass of interstellar gas, but in various components. Following recombination, \LambdaCDM predicts the formation of \DM halos with dense cores and diffuse boundaries \citep{white_1978,blumenthal_1984}. \WGL* surveys confirm this and that galaxies preferentially exist in the cores of \DM halos \citep{mandelbaum_2006,Ingoglia_2022}. We model the \DM halo distribution for a region of the universe comparable in size to the \CCAT survey volume using the \pkp algorithm \citep{Stein_2018,bond_and_myers_1,bond_and_myers_2,bond_and_myers_3}. The \ws simulations \citep{WebSky_paper} then generate \LIM mocks of future \CCAT Observatory surveys. Adjusting the \ws response functions that translate \DM to \LIM observables allows one to study the parameter space of line emission models, and the \pkp approach naturally allows for variation of the early-universe fields to study the parameter space of inflation. 

In Section \ref{sec:theory}, we discuss \LIM surveys and the theory of novel tracers of inflation. Next, in Section \ref{sec:sims}, we summarize the \pkp and \ws simulations used in this work and discuss updates to the \ws \LIM model. In Section \ref{sec:power spectrum}, we discuss constraints on the parameter space of inflation models from the matter power spectrum. Then, in Section \ref{sec:fields}, we consider constraints from halo catalogues generated using \pkp simulations with field-level \png.\footnote{By ``field-level'' \png, we mean that \nong primordial fields are realized in simulations as non-linear functionals of (typically Gaussian) source fields. These can be used by the \pkp algorithm to produce halo catalogues, thus propagating \png to \LSS regime without neccessarily having to take a perturbative approach.} Finally, in Section \ref{sec:cii mocks}, we study the ways in which \ws \cii \LIM mocks can help refine models of cosmological line emission from distant galaxies, and show that these simulations can also be used as a probe of the inflationary parameter space.

\section{Theory}\label{sec:theory}

In this section we outline the theory used in this work. First, we discuss the origins of cosmic structure and tracers of primordial structure formation. Next, we introduce \ping, a novel form of \nong that arises during multi-field inflation. Finally, we discuss the theory of \LIM with the \cii transition line.

\subsection{The origins of cosmic structure and primordial non-Gaussianity}\label{subsec:nonG}

One of the most compelling arguments for inflation is its ability to explain the origin of the inhomogeneities in the \CMB. These fluctuations are seeded by the cosmological scalar field $\zeta(\mathbf{x},t)$, sometimes called a ``comoving curvature perturbation'' because of its resemblance to curvature under the right choice of gauge and slicing. We use the definition on hyper-surfaces of uniform Hubble parameter, $H$, where $\zeta$ is the local deviation from a uniform scale factor \cite{Salopek_1990,Sasaki_1996},
\begin{equation}
    \label{eq: zeta definition on uniform H hypersurfaces}
    \zeta(\mathbf{x},H) \equiv \alpha(\mathbf{x},H)-\bar{\alpha}(H).
\end{equation}
Here, $\alpha$ is the natural logarithm of a scale factor of the sort introduced by \cite{Friedmann_1922}, but with local variations in comoving position, $\mathbf{x}$, $\alpha(\mathbf{x},H) \equiv \ln [ a(\mathbf{x},H)]$. The over bar denotes the ensemble average over realizations of the statistically homogeneous and isotropic field, $\alpha(\mathbf{x},H)$, which is equivalent to a spatial average over a sufficiently large volume per the Ergotic theorem, $\bar{\alpha}(H) \equiv \langle \alpha(\mathbf{x},H)\rangle$ \citep{bbks}. There is a smooth transition in the time evolution between sub-horizon modes $\tilde{\zeta}(\mathbf{k}|k>aH)$, which have time-evolving fluctuations, and super-horizon modes $\tilde{\zeta}(\mathbf{k}|k \ll aH)$, which do not evolve with time \cite{Bardeen_1983,Salopek_1990}. Thus, as inflation causes the comoving horizon scale to shrink, modes that once evolved freeze out. After inflation, the comoving horizon scale increases, and modes $\tilde{\zeta}(\mathbf{k})$ gradually re-enter the horizon, interacting with observable cosmological fields and bringing with them memory of the state of the universe when they exited the horizon. A corresponding response in the matter field is encompassed in a transfer function \cite{bbks},
\begin{equation}
    \label{eq: zeta to delta transfer function}
    \bar{\rho}_m(t)\tilde{\delta}(\mathbf{k},t) = T_{\zeta\to\delta}(k;t,t') \tilde{\zeta}(\mathbf{k},t'),
\end{equation}
where $t$ is whichever time we are interested in looking at the matter field (in this work we use the present $t=0$), and $t'$ is a time after the end of inflation but before \NL gravitational effects introduce additional \nong in the $\zeta$ field. The matter density, $\rho_m$, which represents the density of all forms of energy that cluster under the effects of gravity (so in the \LambdaCDM model, matter and \DM), has an ensemble average (or spatial average over large enough areas) of $\bar{\rho}_m$. Finally, the matter overdensity field, $\delta$, is defined as a function of time and comoving position,
\begin{equation} \label{eq:overdensity}
    \delta(\mathbf{x},t) \equiv \frac{ \rho_m(\mathbf{x},t) - \bar{\rho}_m(t) }{ \bar{\rho}_m(t) }.
\end{equation}

Observations of the \CMB show that the universe at the time of last scattering (and thus $\zeta$ over scales less than the comoving particle horizon at last scattering) was indeed very nearly Gaussian \citep{planckcollaboration_2018_VII,planckcollaboration_2018_IX}. However all inflation models predict some deviations from Gaussian statistics. For the purposes of this work, we express this as a perturbation from a Gaussian field $\zeta_g$,
\begin{equation}
    \label{eq: nG zeta}
    \zeta(\mathbf{x}) \simeq \zetag(\mathbf{x},H) +  \mathcal{O}[\zeta_g^2(\mathbf{x},H)].
\end{equation}
The \CMB and theory further suggest that the dimensionless power spectrum\footnote{Note that in this work, we denote the power spectrum of a field $f$ as $P_{ff}(k)=\langle|\tilde{f}|^2\rangle(k)$, and the dimensionless form as $\mathcal{P}(k) = \frac{k^3}{2\pi^2}P_{ff}(k)$.} of $\zeta$ is nearly scale invariant with an amplitude $A_s$ at a pivot scale, $k_0$, and spectral index $n_s$ nearly 1 \citep{planckcollaboration_2018_VI}.\footnote{In this work, for consistency, we use the convention of \cite{planckcollaboration_2018_VI} and set $k_0=0.05~\mathrm{Mpc}^{-1}$, and assume $A_s$, $n_s$, and \LambdaCDM cosmological parameters have values shown in Table 2 of \cite{planckcollaboration_2018_VI} for TT,TE,EE+lowE+lensing unless otherwise stated.} We therefore realize the Gaussian field $\zeta_g$ in \LSS simulations with this power,
\begin{equation}
    \label{eq:zeta power}
    \mathcal{P}_{\zeta_g\zeta_g}(k) = A_s \left(\frac{k}{k_0}\right)^{n_s-1}.
\end{equation}
A nonlinear functional operating on a \GRF results in a \nong field. Thus, to lowest order, one might construct an \EFT for inflation with a single field $\zetag$ and a non-Gaussian perturbation of the form (\ref{eq: nG zeta}) has a lowest order contribution of the form $\fNLlocal \left[ \zetag^2(\mathbf{x},H) - \langle\zetag^2(\mathbf{x},H) \rangle \right]$. Where we have assumed a constant coefficient,\footnote{$\fNLlocal$ is often expressed in terms of the gravitational potential $\Phi$ rather than $\zeta$, which is related but not identical to this definition.} $\fNLlocal$, and the ensemble average of the square is subtracted to maintain zero mean. As we show in section \ref{subsubsec:pinG}, it must be stressed that this perturbative approach is not the only valid way that \nong can appear in $\zeta$.

Whereas the statistics of the Gaussian field, $\zeta_g$, are fully described by its power spectrum, non-Gaussian fields have non-zero higher-order connected-cumulants. With this simple formulation, $\fNLlocal$ appears in a series expansion of the cosmological three-point function, or bispectrum \citep{Maldacena_2003,Dalal_2008}. The bispectrum\footnote{The bispectrum for a field $f$ can be expressed as $\langle\tilde{f}(\mathbf{k}_1)\tilde{f}(\mathbf{k}_2)\tilde{f}(\mathbf{k}_3)\rangle = (2\pi)^3 \delta_D(\mathbf{k}_1+\mathbf{k}_2+\mathbf{k}_3)B_f(k_1,k_2,k_3)$.} can be peaked in a particular triangular configuration of points with separations representing characteristic scales. These include ``squeezed'' ($k_1 \ll k_2 \approx k_3$), ``equilateral'' ($k_1 \approx k_2 \approx k_3$), ``flattened/folded'' ($k_1 \approx k_2 \approx \frac{1}{2} k_3$), and ``orthogonal'' (which is a combination of equilateral and flattened/folded). For $\fNLlocal$-type \nong, the squeezed configuration is dominant \citep{komatsu_2001}. Certain triangle configurations have been been found to be characteristic of certain inflationary scenarios as discussed in \citep{planckcollaboration_2018_IX} and references therein: single-field inflation models have dominant contributions from equilateral and orthogonal configurations, and $\fNLlocal$ and the squeezed configuration are characteristic of multi-field inflation models. The amplitude of each of the bispectrum configurations discussed here has been constrained from measurements of the \CMB bispectrum to be consistent with zero and confined to between about $\pm1$ and $\pm10$ depending on the configuration \citep{planckcollaboration_2018_IX}. $\zeta$ fluctuations have standard deviation of order $\langle\zeta_g^2\rangle\sim10^{-5}$, so these constraints require a very small non-Gaussian perturbation as, for example, $\fNLlocal$ of 1 would require a non-Gaussian perturbation to have a magnitude less than about $10^{-5}$ times the magnitude of $\zeta_g$. As we will discuss later, if the \nong in $\zeta$ is not correlated with $\zeta_g$, this constraint is significantly relaxed. Because the exact effective potential of inflation is not known, the precise form of \png is not known, nor is the expected magnitude of $\fNL$---it may exist but be arbitrary orders of magnitude smaller than we can measure.

\subsection{Primordial Intermittent Non-Gaussianity (PING)}\label{subsubsec:pinG}

Early inflation models focused on a single field, known as the inflaton, $\phi$, transitioning from a false vacuum state with high potential energy to a true vacuum \cite{Starobinsky_1980,guth_1981,Linde_1982}. If the first two derivatives of the potential are sufficiently small relative to the initial high potential---the so-called ``slow roll'' condition---a phase of accelerated expansion occurs, which we call inflation. Inflation ends with reheating, when the inflaton couples to fields of the standard model of particle physics, converting its kinetic energy (gained by rolling down the potential) into radiation \cite{Albrecht_1982,Linde_1983}. These minimal single-field models are compelling, as they propose solutions to a number of open questions in cosmology, but they also introduce a number of fine-tuning problems \cite{Lyth_1999,Guth_2007,Baumann_2010,Brandenberger_2013}.

There is no necessity that there be only one driving field during inflation; indeed in a given high-energy physical system, there are often multiple relevant fields. We highlight a novel form of \png introduced by \citeMBB, which can be generated during multi-field inflation as a response to short-lived symmetry breaking in the potential of inflationary fields. This particular \nong template is attractive because it is generated as a response to a specific potential feature, regardless of the details of the inflation model that includes this potential feature.

The symmetry breaking can be modelled as a feature in the potential of two inflationary fields. The first, $\phi$, is parallel to the mean flow of the effective inflationary potential during the instability and we therefore refer to it as an inflaton-like field. The second, $\chi$, is perpendicular to the mean flow and is thus referred to as the ``transverse'' field. \nong is generated if the potential is short-lived and has a negative effective mass in the transverse field. The precise mathematical form of the potential has little qualitative impact on the form of the resulting \png if the above are true \cite{morrison2024a}. For the purposes of this work, we model the symmetry breaking in the fashion of \citeMBB as a quartic in the transverse direction with coupling constant $\lambda_\chi$ modulating the overall amplitude and minima at $\chi=\pm v$, known as the vacuum expectation values. The condition for symmetry breaking in the transverse direction is the presence of a negative effective mass term. This occurs if the free-field mass of the transverse field $m_\chi$ is less than a mass-like term that arises in the potential $m_\lambda=\sqrt{\lambda_\chi}v$. We focus on the ``strong'' instability regime, meaning trajectories through field space do not reach the vacuum expectation value before the instability ends. For a detailed discussion of the potential used in this work, see Appendix \ref{app: ping potential}. 

The \nong that is generated is called \ping because the instability is intermittent, turning on at a time $H_i$, reaching a maximum at $H_p$ and turning off at $H_e$. The \nong effects are furthermore localized in position space and in wavenumber space, giving rise to spatial intermittency and scale intermittency. The latter results in the formation of a feature in the power spectrum at a characteristic wavenumber $\kpulse$. It should be stressed, however, that the power spectrum alone does not fully describe any form of \nong, so any statements about features in the \ping power spectrum must be taken with a grain of salt.

For a detailed analysis of \ping and a discussion of regimes beyond the strong instability, which can result in domain walls forming, we direct the reader to the initial work \citeMBB. We will spend the rest of this section discussing how \ping can be implemented in cosmological simulations at the field-level, preserving an approximation of the \nong effects.

\citeMBB demonstrate a component separation procedure that expresses $\zeta$ after inflation as a sum of a largely Gaussian component, $\zeta_g$, and a highly \nong response, $\dzeta$. It must be understood that this component-separation $\zeta_g$ and $\dzeta$ are not the same as the perturbation-theory $\zeta_g$ and the quadratic perturbation of (\ref{eq: nG zeta}). We use similar notation because, in the simulations used in this work, we implement both $\zeta_g$ fields as purely Gaussian fields with power (\ref{eq:zeta power}). The precise mathematical form of the \ping response, $\Delta\zeta$, is derived from component separation of any inflationary fields that interact to produce $\zeta$ fluctuations, and is therefore dependent on inflation model. In this work, the \ping response has a functional form, $\dzeta \simeq F_\mathrm{PING}[\chi(\mathbf{x})]$, which is \nl, and involving multiple transfer functions. The transverse field, $\chi$, is approximately Gaussian, and can therefore be realized in simulations in similar fashion to $\zeta_g$, not requiring lattice simulations for each realization of \pkp halos.

While \ping may leave some measurable effects in certain bispectrum templates, attempting to measure $2$-point functions on the field may be an inefficient summary statistic of \ping. This is because the fluctuations in the transverse field, $\chi$, which sources the \nong, are not strongly correlated with the fluctuations in the inflaton-like field, $\phi$, which sources the Gaussian fluctuations. When \nong and Gaussian fluctuations are sourced by correlated fields, peaks in the \nong field overlap with Gaussian peaks and are amplified. In this case however, there is no such amplification, which can bury the \nong signal, making it more difficult to resolve in measurements of the field. Because \ping also produces spatial intermittency, with the response, $\dzeta$, localized in peaks of characteristic size, we are therefore motivated to look for enhancement of \LSS on corresponding scales instead.


In this work we realize this $\Delta\zeta$ response field in \LSS simulations following a prescription based on the component separation approach by \citeMBB with some simplifications that will be discussed. \citeMBB find that only one time slicing at the transition out of the instability $t_e$ is needed for a good approximation to the \ping response, $\dzeta$. There is a two-step transfer function: first, the potential feature induces a response in the inflaton-like field at $H_e$, $\Delta\phi_e$, which was found to be well approximated by a transfer function from a smoothed quadratic of the transverse field at $H_e$, $\chi_e$,
\begin{equation} \label{eq: ping delta phi}
    \Delta\tilde{\phi}(\mathbf{k},H_e) \simeq T_{W\chi_e^2\to\Delta\phi_e}(k) \mathcal{F}\left[ \big\{ W(\mathbf{x}) * \chi(\mathbf{x},H_e) \big\}^2\right](\mathbf{k}).
\end{equation}
The $\chi$ field is smoothed to avoid numerical effects 
in lattice simulations. Physically, this smoothing must pass at least the comoving wavenumber of the instability band, which has a maximum at $H_p$, 
$k_{\mathrm{inst},p}=(m_\lambda^2-m_\chi^2)^{1/2}$, so we approximate it as a top-hat, $\tilde{W}(k) \simeq \Theta(k_{\mathrm{inst},p}-k)$. See Appendix \ref{app: ping potential} for a discussion of the instability band. \citeMBB show the transfer function, $T_{W\chi_e^2\to\Delta\phi_e}(k)$, to be approximately constant up to $k_{\mathrm{inst},p}$, then approaching zero as $k$ goes to infinity. This was found to be well approximated by a top-hat function, placing the cut-off at $k_{\mathrm{inst},p}$ or $2k_{\mathrm{inst},p}$ were found to result in similar percent level deviations from measurements made from the lattice. $T_{W\chi_e^2\to\Delta\phi_e}(k)$ has an approximate amplitude determined by ballistics or the Hamilton-Jacobi method of inflation \cite{Salopek_1990,Salopek_1991}. Rather than measure the $\chi_e$ power spectrum directly from lattice simulations for every realization used in \LSS forecasts, we use a prototypical power that is calculated in linear theory and rescale its amplitude and characteristic scale,\footnote{Early-universe simulations use a scale factor set to $a^\mathrm{EU}_i=1$ at the beginning of the instability. We convert to the \LSS units that set $a_0^\mathrm{LSS}=1$ at the present by multiplying by factors of $a^\mathrm{LSS}/a^\mathrm{EU}$.} $a_e$, to match the power spectrum measured from the lattice. The amplitudes of the $\chi_e$ power and $T_{W\chi_e^2\to\Delta\phi_e}(k)$ are absorbed into a single fudge factor $\mathcal{Q}$, which is used to control the amplitude of the \ping fluctuations. The smoothing is then implemented with unit amplitude, and transfer function is thus implemented as a low-pass filter, $T_{W\chi_e^2\to\Delta\phi_e}(k) \simeq\mathcal{Q}^2\Theta(2k_{\mathrm{inst},p}-k)$, up to factors of the scale factor. \citeMBB found that similar approximations result in a difference of only a few percent in the correlations of fields compared with those measured from the lattice simulations in the strong instability regime.

The perturbation, $\Delta\phi$, at $H_e$ is related by a transfer function to the \ping response in its frozen-in asymptotic state, which we denote $\dzeta_f$ \cite{morrison2024a}, 
\begin{equation} \label{eq: ping delta zeta}
    \dtildezeta_f(\mathbf{k}) \simeq T_{\Delta\phi_e \to \dzeta_{f}}(k) \Delta\tilde{\phi}(\mathbf{k},H_e).
\end{equation}
The final scalar field $\zeta\simeq\zeta_g+\Delta\zeta_f$ is related to the matter field by the transfer function (\ref{eq: zeta to delta transfer function}), which we assume to be unchanged from the primordially Gaussian form at least in the strong instability regime.
\begin{figure}
    \vspace{-2em}
    \centering
    \includegraphics[width=0.67\linewidth]{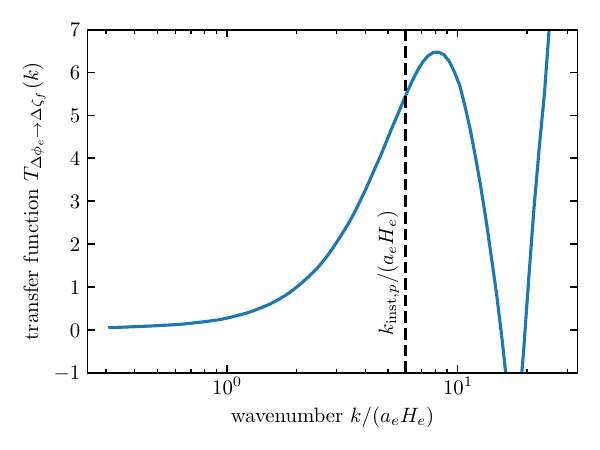}
    \vspace{-1em}
    \caption{The transfer function from from the inflaton-like field, $\Delta\phi_e$, to the asymptotic freeze out \ping $\zeta$ response. The wavenumber, $k$, is expressed in units of the comoving Hubble scale at the end of the instability, $a_eH_e$. Modes, $\Delta\tilde{\zeta}_f$, with $k$ larger than about $k_{\mathrm{inst},p}/(a_eH_e)$, shown as the black dashed line, are subject to the $\chi$ smoothing and $T_{W\chi^2_e\to\Delta\phi_e}(k)$ transfer function which are low-pass filters, so high-$k$ behaviour of this transfer function will not affect $\zeta$.}
    \label{fig:Delta phi to Delta zeta transfer func}
\end{figure}
The transfer function, $T_{\Delta\phi_e \to \dzeta_f}(k)$, has a more complicated $k$-dependence and can be measured from early-universe simulations. This transfer function is shown in Figure \ref{fig:Delta phi to Delta zeta transfer func}. The wavenumber is expressed in terms of the comoving Hubble scale at the end of the instability, $a_eH_e$. The \ping response in the inflaton-like field, $\Delta\phi_e$, is derived from the transfer function, $T_{W\chi^2_e\to\Delta\phi_e}(k)$. Both the smoothing kernel, $W$, and the transfer function, $T_{W\chi^2_e\to\Delta\phi_e}(k)$, are low pass filters, so high-$k$ modes of $\Delta\phi_e$ are zeroed. Consequently, the negative portion of the transfer function, $T_{\Delta\phi_e\to\Delta\zeta_f}<0$, and its behaviour at high $k$, do not affect the response, $\Delta\zeta_f$, in the strong instability regime considered in this work.


In the strong instability regime that we are working under, the \ping response introduces an enhancement of the power spectrum\footnote{Note that equation \ref{eq:PING power spectrum feature} can equivalently be expressed in terms of the matter power (\ref{eq:matter power}) shown in Figure \ref{fig:PING vs CMB matter power} by replacing the $\zeta$ powers with corresponding matter power.} that is localized in $k$ with a characteristic shape that can be compared to the nearly scale-invariant Gaussian $\zeta$ power (\ref{eq:zeta power}),
\begin{equation}
    \label{eq:PING power spectrum feature}
    \mathcal{D}(k) = \frac{ \mathcal{P}_{\zeta\zeta}(k) }{ \mathcal{P}_{\zeta_g\zeta_g}(k) }-1.
\end{equation}
The enhancement has a maximum $\mathcal{D}_0 \equiv \max \mathcal{D}(k) = \mathcal{D}(k_\mathrm{pulse})$, and asymptotes to zero at wavenumber, $k_\mathrm{cutoff}\simeq k_{\mathrm{inst},p}$. In this work, we focus on the regime in which only a few $e$-folds of expansion in the scale factor, $a$, occur during the instability, so $k_\mathrm{cutoff}$ is of order a few times $k_\mathrm{pulse}$. Because we are agnostic to the exact form of the \EFT for inflation outside the instability, the characteristic scale, $a_e$, and the amplification factor, $\mathcal{Q}$, are treated as free parameters in \LSS simulations, we will demonstrate constraints on them in later sections. The more physically meaningful control parameters, the characteristic wavenumber, $\kpulse$, and power spectrum enhancement amplitude, $\mathcal{D}_0$, can be calculated from $a_e$ and $\mathcal{Q}$, or measured approximately from the \pkp initial conditions fields for each \ping realization.

There have been many efforts to constrain $\fNL$-type \png with cosmological observables \citep{Slosar_2008,Leistedt_2014,planckcollaboration_2018_IX,Martinelli_2021,mueller_2022,chen_2025}, including with \LIM \citep{Camera_2013,moradinezhad_2019,Karagiannis_2020,Liu_2021,Kopana_2024}. While multi-field inflation models that give rise to \ping may also have non-zero $\fNLlocal$, it could be suppressed as discussed earlier because $\zeta_g$ and $\dzeta$ are uncorrelated. Therefore, finding other methods to measure \ping could lead to new insights into inflation. Furthermore, \nong fields in principle depend on an infinite number of higher connected cumulants, and there is no guarantee that just the bispectrum will be especially informative. For this reason, in this work we use ``field-level'' \png, meaning that \nong is implemented through theory-informed functionals of inflationary fields as described above. Thus all higher cumulants exist in realizations of cosmic fields (at least up to the approximations that we have made herein), and can be propagated to \LSS observables such as the \HMF and mock sky maps.

\ping may be considered to be of a class of \png models which introduce distinct features in the power spectrum. It is therefore important to distinguish the difference between a Gaussian field with a power spectrum feature from a non-Gaussian field with a similar feature. Spike power spectrum models have been studied in the literature with features generated by various mechanisms \citep{Kofman_1988,Salopek_1989,Salopek_1990,Salopek_1991,Bond_2009,Junaid_2018,mylova_2022}. Many of these models may benefit from similar analysis to that carried out below. 

\subsection{Cosmology with Line Intensity Mapping (LIM)}\label{subsec:lim}

\LIM is an emerging technique to directly measure the \LSS of the universe during the \EoR. Photons emitted from the interstellar and intergalactic gas emit at a number of characteristic electronic, atomic and molecular lines. The redshifting of emission lines due to the Hubble flow allows for tomographic measurements of the cosmological \LIM signal when measured spectroscopically \cite{Scott_1990,Righi_2008,Gong_2011}. Using wide-field surveys, the cosmological \LIM signal can be measured as a bulk property rather than having to resolve individual galaxies, and can thus be matched to \LSS fluctuations and used to reconstruct a three-dimensional map of the cosmic structure during the \EoR.

While other tracers of the \LSS during the \EoR exist, most trace extremely bright sources, or are projected effects sourced mostly by structures at relatively low redshifts. For example, surveys of \BAO catalogue high-redshift galaxies and quasars, but these bright sources are biased toward the most dense cluster regions, thus excluding information about the rest of the cosmic web \citep{SDSS-IV_2021,DESI_2025,bechtol_2025,DES_Collaboration_2025}. Conversely, photons can be measured from dusty star-forming regions that glow in the infrared to produce a \CIB \citep{Partridge_1967,planckcollaboration_XVIII}, however this is a projected background, integrated along the line of sight not measured as a function of redshift, and its strongest contribution is sourced when the star formation rate peaked during Cosmic Noon ($z\sim2$). In contrast, the cosmological \LIM signal is fully three-dimensional, meaning the intensity can be matched directly to \LSS evolving with redshift. Similarly, \LSS between us and the surface of last scattering leaves characteristic traces in the \CMB signal: Compton scattering by hot gas or bulk flows of gas comprise respectively the thermal and kinematic (or kinetic) \SZ effects \cite{Sunyaev_1970,Sunyaev_1972,Sunyaev_1980,planckcollaboration_2015_XXII,planckcollaboration_2015_XXVII,planckcollaboration_2015_XXXVII}, gravitational redshift by \CMB photons passing through growing overdense regions produces the \ISW effect \cite{sachs_wolfe_1967,planckcollaboration_2015_XXI}, and \WGL magnifies and deflects \CMB photons around massive structures \cite{planckcollaboration_2018_VIII}. Once again however, each of these effects is projected along the line of sight and peaks at redshifts $z\lesssim5$.

\LIM provides a view of the higher redshift universe during the \EoR, but it is challenging to observe, as cosmological line emission must be distinguished from relatively bright and ubiquitous galactic foregrounds. In much the same way that \CMB foregrounds have been validated on cosmological simulations such as \ws \cite{WebSky_paper,Coulton_2024,ACT_2025b}, sophisticated simulations of cosmological line emission have the potential to validate \LIM models and distinguish signal from foregrounds. In the following section, we discuss the \pkp and \ws simulations, which we use to generate the \LIM mock maps. These maps are statistically analogous to observations, thus we use them to make forecasts for models of the cosmological line emission signal and for cosmological parameters such as \png.

Galactic and instrumental foregrounds will be largely independent for surveys of separate cosmological emission lines or other tracers of \LSS, so any correlations between surveys can be mapped to cosmological structure. A robust mock pipeline could further support such cross-correlation studies. \pkp and \ws are ideal candidates as they already include treatments of the \CIB, \SZ effects, and \WGL along with several line emission models.\footnote{These mocks as well as a linear treatment of the \ISW effect are publicly available at \href{https://uoft.me/websky}{uoft.me/websky}.} Correlations between \LIM and other \LSS tracers ($\mathrm{\LIM}\times\mathrm{CMB~foregrounds}$, $\mathrm{\LIM}\times\mathrm{\BAO}$) can validate which models of the cosmological \LIM signal best capture the cosmological structure, and correlations between distinct emission lines ($\mathrm{\LIM}\times\mathrm{\LIM}$) will unveil the \LSS of the cosmic web deeper into the \EoR than has thus far been observed.

In this work, we focus on mocks for the \CCAT EoR-Spec instrument, which will measure \cii line emission in a redshift range $z\in[3.5,8]$ \citep{Freundt_2024}. The \cii line is emitted by the $J=3/2 \to 1/2$ fine structure transition of singly ionized carbon, $\mathrm{C}^+$, at a rest wavelength, $\lambdacii = 157.74~\mathrm{\mu m}$ (or frequency, $\nucii \simeq 1900.548 ~ \mathrm{GHz}$) \citep{lagache_2018}. Much of the cosmological \LIM signal for these mm-wave observations is emitted from photodissociation regions (PDR) around bright UV photon sources, such as early stars \cite{lagache_2018}. The PDR is far enough from the source that the flux is insufficient to fully ionize hydrogen, but where the interstellar gas is still dense enough for a significant \LIM signal from $\mathrm{C}^+$ and neutral oxygen, $\mathrm{O}$. It should be noted that some models of the cosmological \cii signal have a significant contribution not sourced from PDR, including the model used in this work, \cite{Liang_2023}, which is based on the Feedback In Realistic Environments (FIRE) simulations \cite{Feldmann_2016,Feldmann_2017,Angeles-Alcazar_2017,Feldmann_2023}. Model discrepancies such as these further motivate the need for comparative surveys of cosmological line emission models. Such studies could be built using the forecast ansatz laid out in this work.

For a more complete theory overview of \LIM, see \cite{bernal_2022}. For a review of some of the instruments in the \LIM landscape, see \cite{kovetz_2017}.

\section{Simulations}\label{sec:sims}

In this section, we summarize the simulations used in this work. First, we provide an overview of the \pkp simulations, which are used to generate realizations of the distribution of \DM in cosmological volumes. Then, we summarize the \ws simulations, which use the results of the \pkp calculation to realize mock maps of the sky in a suite of observables. The \ws \cii mocks and \pkp \DM halo catalogues are publicly available at \href{https://uoft.me/webskycii}{uoft.me/webskycii}.

\subsection{The \textup{\pkp} dark matter halo simulations}\label{subsec:peak patch}

After recombination, the \NL effects of gravity begin to introduce non-Gaussian modes into the primordial, very-nearly-Gaussian matter field. Evolving these fields all the way to the present requires a numerical rather than analytic treatment because the field becomes highly non-Gaussian in dense cluster regions. Any choice of numerical methods will introduce some inaccuracies, \nbody and \hydro simulations for instance introduce some minimum mass resolution and will typically not have a fully general relativistic treatment of dynamics. Therefore, when choosing a simulation, one is naturally faced with a question of which assumptions achieve maximum efficiency for a desired level of accuracy in the final simulation products. For large cosmological simulations, the evolution with redshift must also be accurately treated. Such catalogues that take into account light travel time are called ``light cone'' catalogues.

We employ the \pkp algorithm \citep{Stein_2018,bond_and_myers_1,bond_and_myers_2,bond_and_myers_3} to simulate the distribution of \DM halos in a region of the universe comparable to the CCAT survey volume. \DM halos are diffuse and do not have an obvious boundary. In this work, we use the ``$M_{200m}$'' definition, meaning the radius of a halo is defined such that the mean halo density is $\bar{\rho}_\mathrm{halo} = \Delta\bar{\rho}_m$ where $\Delta=200$, which is somewhat larger than that of the virial density contrast in Einstein-de Sitter space, $\Delta_\mathrm{vir}\simeq178$ \cite{Gunn_Gott_1972}. \pkp locates candidate halos by smoothing an initial, linear-theory matter field with a series of top-hat smoothing radii, $R_{\mathrm{TH},i}$, then models the collapse of \DM halos as overdense ellipsoidal regions with homogeneous mass density and initial radius, $R_\mathrm{halo} = R_{\mathrm{TH},i}$. Each halo is collapsed according to the strain of the matter field and if it reaches the critical density $200\bar{\rho}_m$, it is included in the catalogue. Finally, a merging and exclusion routine is run to ensure mass is not double counted, and halos are propagated to final positions using $2^\mathrm{nd}$ order Lagrangian perturbation theory. This is possible because all the hot virialized material lies within halos, and the medium through which they move is underdense and governed by weakly \NL dynamics.

There is natural support for light cone catalogues as any halo that does not collapse by the redshift at which it is observed is left out of the final catalogue. This is an advantage over large \nbody or \hydro simulations that have to break up a volume into shells of redshift, introducing an unnatural binning effect. Because peaks evolve from a matter field, \png and specifically \ping can easily be implemented fully at field-level. The simulations are extremely efficient compared to \nbody or \hydro simulations because each region of the universe is modelled only up the level required to locate halos. The space between halos is modelled with low-order perturbation theory, and the halo interiors are not treated (see Section \ref{subsec:CII I model}). Because of its efficiency, many \pkp realizations with varied cosmologies can be made for comparative study. Realizations of the dominant Gaussian field, $\zetag$, are produced by convolving a Gaussian white noise field with the two-point correlation function predicted by \LambdaCDM that is calculated for a given cosmology using the CLASS Boltzmann solver \citep{CLASS_2011}. For our fiducial model, we use the \emph{Planck} Collaboration 2018 results for cosmological parameters \cite{planckcollaboration_2018_VI} (see table 2, TT,TE,EE+lowE+lensing). Realizations of $\dzeta$ are generated by applying the transfer functions (\ref{eq: ping delta zeta},\ref{eq: ping delta phi}) to the field, $\chi$, which is generated analogously to $\zetag$ using the $\chi$ correlation function from lattice simulations \cite{morrison2024a}. While we focus on features in the power spectrum, in later sections, it should be kept in mind that higher order correlations are in place in all \pkp halo catalogues and \LIM mocks.

\pkp is classified as a semi-analytic \DM halo code because of the treatment of halo collapse based on analytic \HEC and perturbation theory. \pkp has been validated against \nbody codes and has been compared with similar approximate methods for generating mock halo catalogues, performing very competitively \citep{mock_comparison_I,mock_comparison_II,mock_comparison_III}.

Much of this work is based on the \CCAT \DSS, which observes \cii at redshifts $z\in[3.5,8]$ in a $4 ~ \mathrm{deg}^2$ patch of sky. To simulate this volume, a pair of adjacent cubic \pkp light cone catalogues with total comoving volume of $2\times(1100 ~ \mathrm{Mpc})^3$ were created for each cosmological realization. The cubes are centred at comoving distance $7.5$ and $8.6 ~ \mathrm{Gpc}$ from the observer and realized at field-level with a total of $2\times5640^3$ ($2\times5586^3$ excluding buffers) voxels. The \pkp algorithm was run on the SciNet Niagara supercomputer with each half run as as $21^3$ parallel cubic sub-volumes following the parallelization scheme laid out in \cite{Stein_2018}. This gives a fundamental field resolution of $197 ~ \mathrm{kpc}$ and a minimum halo mass of $4.3 \times 10^9 ~ M_\odot$. Using adjacent simulation volumes results in minimal edge effects as was done in the octant treatment of \wsone \citep{WebSky_paper}. For comparison, the \wsone \CMB foreground mocks have a field resolution of about $1.25 ~ \mathrm{Mpc}$ and a minimum halo mass of about $2.8\times10^{10} ~ M_\odot$. The \wsone catalogue has a redshift $z<4.6$, thus the increased resolution the \DSS and future \LIM mocks are necessary to resolve halos out to $z\lesssim8$.

\subsection{The \texorpdfstring{\ws}{WebSky} mock sky map simulations and the cosmological \texorpdfstring{\textup{\cii}}{[CII]} intensity model}\label{subsec:CII I model}

The empirical forward-modelling code \ws is used to transform light-cone \DM halo catalogues into a suite of observables \citep{WebSky_paper}. Mock maps are obtained using response functions that characterize the galaxy-halo connection in a given observable. The term ``response function'' is chosen suggestively as the \DM is a biased tracer of the matter distribution, and perturbations from a uniform \DM field invoke a characteristic response in observable cosmological signals. While a response function to a physical system in classical mechanics such as a harmonic oscillator is linear (taking the form of a Green's function for the differential operator), cosmological observables are highly nonlinear and require more specialized approximate methods. The \ws response functions are based on generalized NFW pressure profiles for the gas in halo interiors \citep{bbps,bbps_ii,bbps_iii,bbps_iv}, halo occupation distributions (HODs) for point sources, and a field effect for signals that have significant contributions coming from beyond the $M_{200m}$ halo cutoff radius.

For \LIM mocks, \DM halos are assumed to be biased tracers of gases in the \ISM that, when excited, emit radiation in characteristic spectral lines. The model for the \cii line emission follows that of \citeHCBL. The \cii luminosity from a given halo of mass $M_\mathrm{halo}$ and redshift $z$ is
\begin{equation}
    \label{eq: CII luminosity per halo}
    \frac{ \Lciihalo(M_\mathrm{halo},z) }{ L_\odot } = \alphacii \frac{ \mhi(M_\mathrm{halo}) }{ M_\odot } \frac{ \zeta_Z Z(M_\mathrm{halo},z) }{ Z_\odot },
\end{equation}
where the over all \cii amplitude is taken to be $\alphacii = 0.024$. The mass of \hi in a halo in this model follows directly from equation 6 of \citeHCBL. The \hi mass is a power law $\mhi(M_\mathrm{halo}<M_\mathrm{min}) \sim M_\mathrm{halo}^{0.74}$, with a cutoff mass $M_\mathrm{min} = 2.0\times10^{10} ~ M_\odot$ above which the \hi mass decays $\mhi(M_\mathrm{halo}>M_\mathrm{min}) \sim e^{ - ( M_\mathrm{min} / M_\mathrm{halo})^{0.35}}$. The metallicity is given a statistical scatter, represented by the factor $\zeta_Z$, drawn from a lognormal distribution,
\begin{equation}
    \label{eq: log-norm Z scatter}
    p(\zeta_Z,\sigma_Z)
    =
    \frac{1}{ \sigma \zeta_Z \sqrt{2\pi} } \exp\left( - \frac{ ( \ln\zeta_Z - \mu)^2 }{ 2\sigma^2 } \right).
\end{equation}
A uniform scatter of points in log-space results in a skewed linear mean, so the logarithmic mean $\mu=-\sigma^2/2$ is chosen to preserve the linear mean value of metallicity, $Z(M_\mathrm{halo},z)$; see appendix B of \cite{chung_2019} for more details. The fiducial value of the natural logarithmic scatter, $\sigma_Z=\sigma/\ln10$, as given in \cite{vizgan_2022}, $\sigma_Z = 0.4 ~ \mathrm{dex}$,
is used in forecasts.

\LIM surveys use spectroscopy to produce three-dimensional, tomographic maps of characteristic line emission at a frequency, $\nu$. An area of the sky, $\Omega$, is divided into pixels of angular size $d\theta\times d\phi$ with uniform frequency bins of width $d\nu$. Each frequency bin can be identified with a redshift range since the Doppler-shifted frequency is $\nu_\mathrm{obs} = (1+z)\nu$. Because comoving distance does not vary linearly with frequency, the voxels in the resulting tomographic map have non-uniform comoving volume, which may introduce complications in interpreting certain summary statistics such as the power spectrum of the tomographic \LIM measurements.

The \cii intensity, $I$, of a mock map with frequency $\nu = \nucii$ is obtained by integrating this luminosity from (\ref{eq: CII luminosity per halo}) for all halos in a given voxel at a position on the sky, $\bm{\hat{\theta}}_\mathrm{voxel}$, and observed frequency, $\nu_\mathrm{obs}$,
\begin{equation}
    \label{eq: CII intensity}
    \Icii(\bm{\hat{\theta}}_\mathrm{voxel},\nu_\mathrm{obs}) = \frac{1}{4\pi\nucii H(z)} \frac{\int_\mathrm{voxel} \Lcii d^3\mathbf{r}}{\int_\mathrm{voxel} d^3\mathbf{r}},
\end{equation}
where the redshift of the voxel is $z = \nu_\mathrm{obs}/\nucii-1$.

\citeHCBL considered only redshifts $z\simin[6,8]$, and assumed a constant luminosity distance, which results in only slight error because the luminosity distance is related to the comoving transverse distance, or the comoving side length of the approximately square voxel in the directions transverse to the line of sight, $\chi_T$, by $D_L \simeq (1+z)\chi_T$. In this work we consider a redshift range $z \in [3,8]$, meaning this approximation becomes less tenable, so we include this factor in our \cii response function.

In this work, we focus on two scenarios for instrumental noise which were also used in \citeHCBL, one based on the planned \DSS by the CCAT Observatory \citep{CCAT_Prime_Collaboration_2022}, and one based on a future \Stwo that could be housed at the CCAT observatory, given that the Prime-Cam instrument occupies only the central 40\% of the available focal plane of the 6-metre \FYST. For simplicity, these two scenarios are referred to as \DSS and \Stwo respectively.

\DSS observes an area of 4 square degrees at frequencies $\nu \in [210 ~ \mathrm{GHz}, 420 \mathrm{GHz}]$ with the \EoRSpec instrument. The \cii line is redshifted to $z\in[3.5,8]$ in this frequency band. As the pixelization for the entire intensity cube, we use the largest Airy disk in this frequency range, which has standard deviation $\theta_\mathrm{FWHM}/\sqrt{8\ln2} \sim 20''$. The survey lasts 2000 hours with 120 feeds for a total of $2.4\times10^5$ spectrometer-hours. 

\Stwo assumes a factor of 20 increase in spectrometer-hours per field. This could be achieved with an updated spectrometer filling the full focal plane of the \FYST, as well as advances in spectrometers that are expected to enable $10^6-10^7$ spectrometer-hour surveys in the next decade \citep{karkare_2022}.

\section{Constraints on \ping-like features from the matter power spectrum}\label{sec:power spectrum}

\begin{figure*}
    \centering
    \vspace{-3em}
    \includegraphics[width=0.8\linewidth]{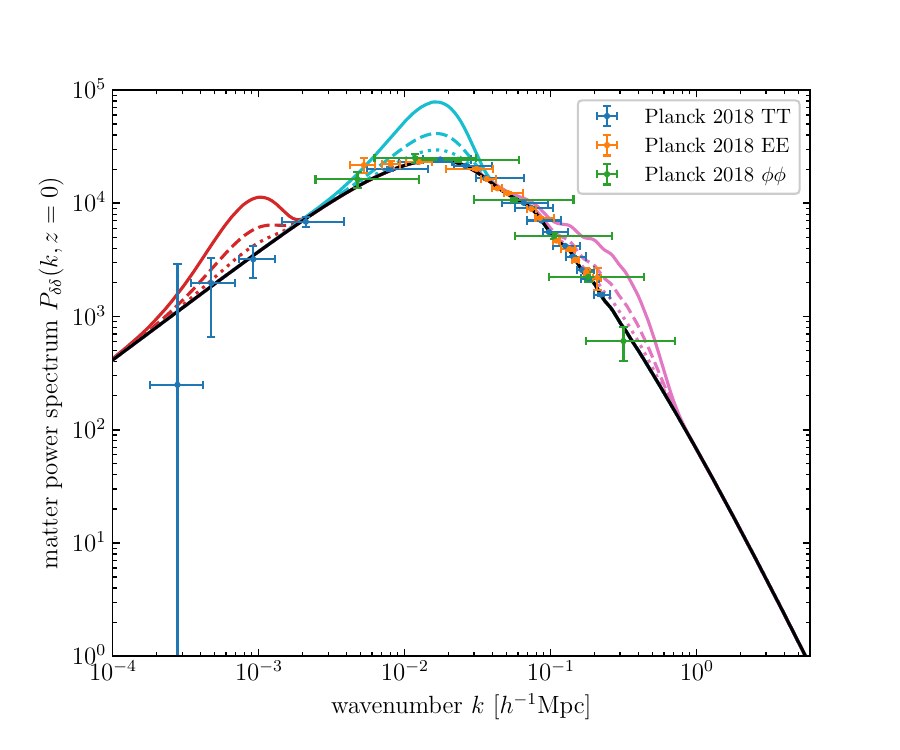}
    \vspace{-1em}
    \caption{
        Possible \ping features at $\kpulse$ of $4\times10^{-4}$, $8\times10^{-3}$, and $1.6\times10^{-1} ~ h/\mathrm{Mpc}$ (in red, cyan and pink respectively), and amplitudes $\mathcal{D}_0$ of $2$, $2\times10^{-1/2}$, and $2\times10^{-1}$ (solid, dashed and dotted lines respectively) contrasted with the 2018 \planck matter power constraints with $1\sigma$ error bars as shown in Figure 19 of \cite{planckcollaboration_2018_I}. The solid black curve shows the primordially Gaussian theory curve with best fit parameters from \planck using the \class Boltzmann solver. Tight constraints on the matter power at the peak of the power spectrum heavily rule out even a small \ping effect, but further from the peak moderate \ping could be possible.
    }
    \label{fig:PING vs CMB matter power}
\end{figure*}
While the \pkp simulations implement full, field-level primordial non-Gaussianity, in this section we explore a \GRF generated with an identical matter power spectrum to that of a universe with \ping. Figure \ref{fig:PING vs CMB matter power} shows a series of potential \ping features in the matter power spectrum, which we define as
\begin{equation} \label{eq:matter power}
    P_{\delta\delta}(k,z) \equiv \left\langle \left| \bar{\rho}_m(z) \tilde{\delta}(\mathbf{k},z) \right|^2 \right\rangle.
\end{equation}
The \planck Collaboration 2018 best fit derived from TT, EE and lensing harmonic power spectra measurements are shown as points with $1\sigma$ error bars \citep{planckcollaboration_2018_I}. The dotted, dashed and solid lines represent moderate, strong and extreme \ping features. The red, cyan and pink curves are chosen to fall respectively in the $P_{\delta\delta}(k)\propto k$ regime, the peak in matter power amplitude, and the transition between \BAO and the $P_{\delta\delta} \propto k^{-3}$ regime. Both the amplitude, $\mathcal{D}_0$, and characteristic wavenumber, $k_\mathrm{pulse}$, of the \ping feature are free parameters in the early-universe model, but it is plain to see that the values they can take are not completely unconstrained. In all cases, the extreme \ping feature cases are clearly disfavoured by $4\sigma$ or more, and where the tightest constraints have been measured around the matter power peak and the \BAO, even the moderate \ping case is heavily disfavoured.

\begin{figure*}
    \centering
    \vspace{-1em}
    \includegraphics[width=1.0\linewidth]{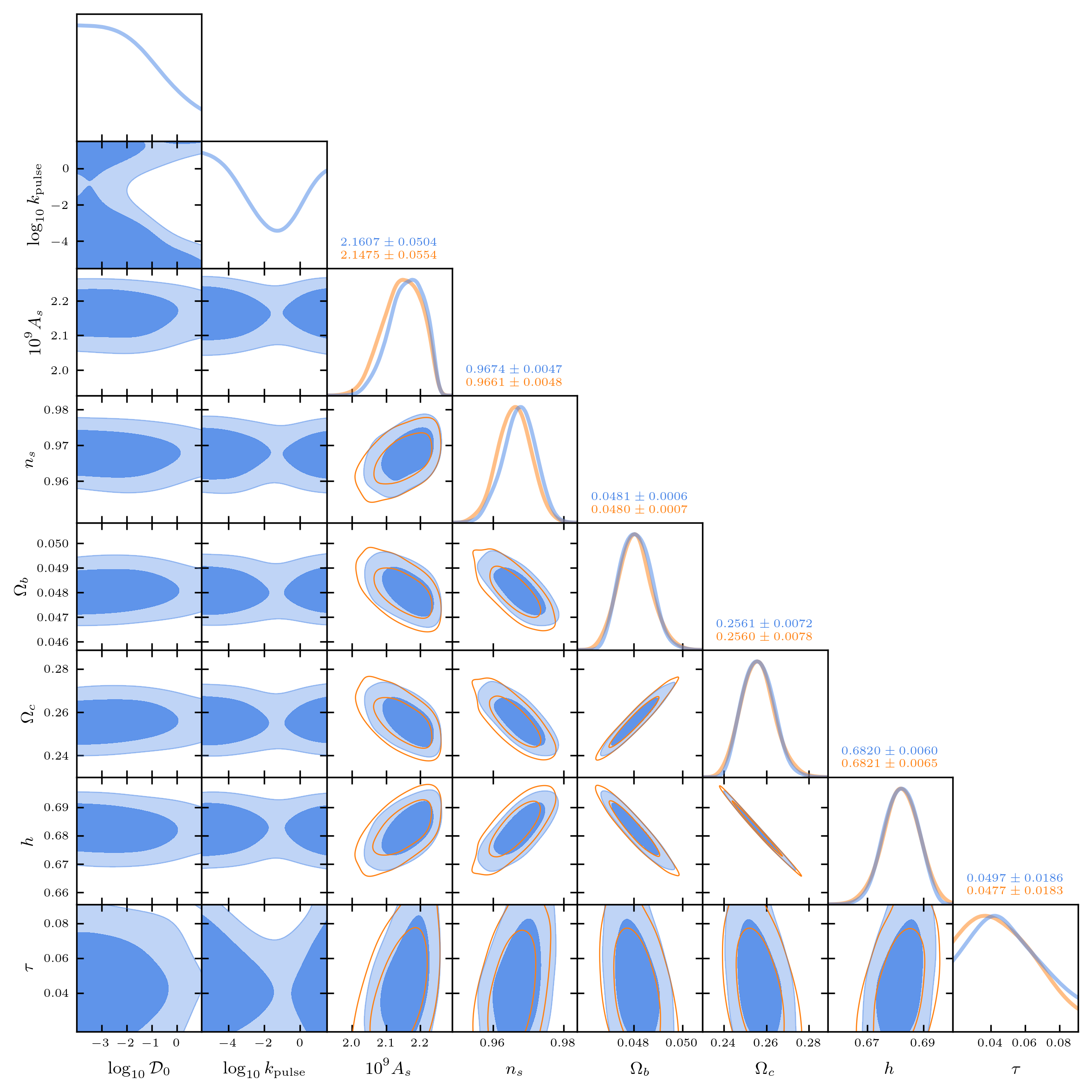}
    \vspace{-1em}
    \caption{
        Posterior probability distributions generated using the \cobaya \MCMC and \class Botlzmann solver codes. The solid blue and light blue shaded regions denote $1\sigma$ and $2\sigma$ joint posterior probability contours for a universe with a \ping-like feature in the primordial power spectrum. Orange contour lines are the $1\sigma$ and $2\sigma$ joint posterior probability contours for a power law primordial power spectrum. The Posterior \PDF{} for each parameter is shown along the diagonal with the mean and $1\sigma$ uncertainties listed above in colours corresponding to the contours. As suggested in Figure \ref{fig:PING vs CMB matter power}, posteriors disfavour $\kpulse$ near the peak of the power spectrum and place an upper bound on the \ping feature amplitude, $\mathcal{D}_0$. \cobaya and \class do not include moments beyond the two-point, so this is in effect a constraint on a primordially Gaussian universe with a \ping-like feature, not a true constraint on \ping.
    }
    \label{fig:CobayaMCMC}
\end{figure*}

For a more rigorous investigation of the constraints, we employ the \cobaya \MCMC code \citep{torrado_2021} with a custom implementation of the \class Boltzmann solver \citep{blas_2011} representing a primordially Gaussian universe with a \ping-like feature in the matter power spectrum. For consistency with \pkp simulations, we again use \planck Collaboration's 2018 results for TT,TE,EE+lowE+lensing $C_\ell$ likelihoods and priors \citep{planckcollaboration_2018_VI} in this analysis. Over the scales considered in this work, the shape (\ref{eq:PING power spectrum feature}) of the \ping feature changes relatively little. For this reason, we simplify the sampling algorithm implemented in the MCMC calculation to fit the \ping feature shape for a single run of the early universe simulations, and then rescale $\mathcal{D}_0$ and shift $\kpulse$ accordingly.

While non-Gaussianity inherently involves higher order correlators, this analysis involves only the power spectrum and therefore assumes primordial fields are Gaussian. Full field-level analysis is carried out in subsequent sections. In both cases though, we make the fundamental assumption that \ping effects are limited to the primordial. \ping produces \nong in the matter field and leaves power spectrum features. These are then assumed to contribute only excess gravity, not any extra terms in the equations of motion. Particularly on scales smaller than galaxies, $\kpulse > 10 ~ \mathrm{Mpc}^{-1}$, some \NL clustering could be expected from strong \ping effects. Similar work has been done to determine the effects of $\fNL$-type \nong on excursion set theory \citep{moradinezhad_2013}; we leave a full treatment of this in the \pkp simulations to future work.

The results, after marginalizing over nuisance parameters, are shown in Figure \ref{fig:CobayaMCMC}. A second \cobaya run with a Gaussian primordial $\zeta$ power is shown with the overlaid orange contour lines, demonstrating that the introduction of \ping features has only a small effect on the joint posteriors for \LambdaCDM parameters. The joint posteriors involving \ping-like power spectrum feature parameters can be understood in analogy to features of mountain ridges, steeply sloped away from best fit values of the \LambdaCDM parameters, but less constrained in the \ping feature parameters, creating a ridge running parallel to the $\kpulse$ and $\mathcal{D}_0$ axes. The amplitude, $\mathcal{D}_0$, is the factor by which the \ping feature $P_{\delta\delta}(\kpulse)$ is greater than the underlying Gaussian matter power $P_{\delta_g\delta_g}(\kpulse)$, so for any $k$ at which the matter power has been measured, the amplitude of the \ping feature is an upper bound, typically $\mathcal{D}_0 \lesssim 1$. As a result, joint posteriors of $\mathcal{D}_0$ with each of the \LambdaCDM model parameters resemble the foot of a mountain ridge, descending gradually to the lowlands as $\mathcal{D}_0$ increases toward $\mathcal{O}(1)$ and falling off steeply as the \LambdaCDM model parameters diverge from their \planck 2018 best fit values. The wavenumber at which the \ping feature occurs, $\kpulse$, reaches a saddle point, not a maximum, around $(10 ~\mathrm{Mpc})^{-1}$. This is the regime in which the tightest constraints have been measured on the power spectrum as we can see in Figure \ref{fig:PING vs CMB matter power}. As a result, we see a saddle in the joint posterior surface of $\kpulse$ and each of the \LambdaCDM model parameters, which is analogous to a mountain col, a broad mountain ridge with a low pass between two peaks. Finally, the combination of these two effects in the joint posterior of $\kpulse$ and $\mathcal{D}_0$ resembles a coulee, a sloped drainage valley that channels water from the highlands (at low $\mathcal{D}_0$ and high or low $\kpulse$) down to the lowlands (at $k$ around the matter power peak and \BAO, or at $\mathcal{D}_0 \gtrsim 1$).

Further constraints could be made at power-spectrum level by incorporating more CMB observations such as SPT \citep{keisler_2011,crites_2015} and ACT \citep{qu_2024,louis_2025}, along with \LSS tracers such as \LRGs from SDSS \citep{oka_2014} and DESI \citep{novellmasot_2025}, Lyman-$\alpha$ forest measurements from BOSS \citep{lee_2013} and cosmic shear with DES \citep{troxel_2018}. As the primary focus of this work is on constraining power of next-generation \LIM surveys, we leave this to future work.

We continue the exploration of the differences between \ping-like features in the matter power of a \GRF and field-level \ping by realizing cosmological fields and \pkp \DM halo catalogues in the next section. With this exercise, we can distinguish between the effects of having a \ping-like spike in the matter power of a \GRF and having the fully non-Gaussian effects of \ping.

\section{Constraints on field-level \ping from \pkp halos}\label{sec:fields}

We have demonstrated constraints on a \ping-like power spectrum feature in a \GRF from measurements of the \CMB. In this section, we discuss why the power spectrum may not be the best tool to constrain effects that produce power spectrum features in non-Gaussian fields. Instead, for the remainder of this work we implement \ping fully at field-level by realizing primordial Gaussian fields and applying non-linear functionals (\ref{eq: ping delta phi},\ref{eq: ping delta zeta}) to produce non-Gaussian fields. These are then used to produce \pkp halo catalogues which can be contrasted with halos generated from a \GRF.

\subsection{Power spectrum features in Gaussian and non-Gaussian fields}\label{subsec:PING features in G and nonG fields}

\begin{figure}
    \centering
    \vspace{-1em}
    \includegraphics[width=1.0\linewidth]{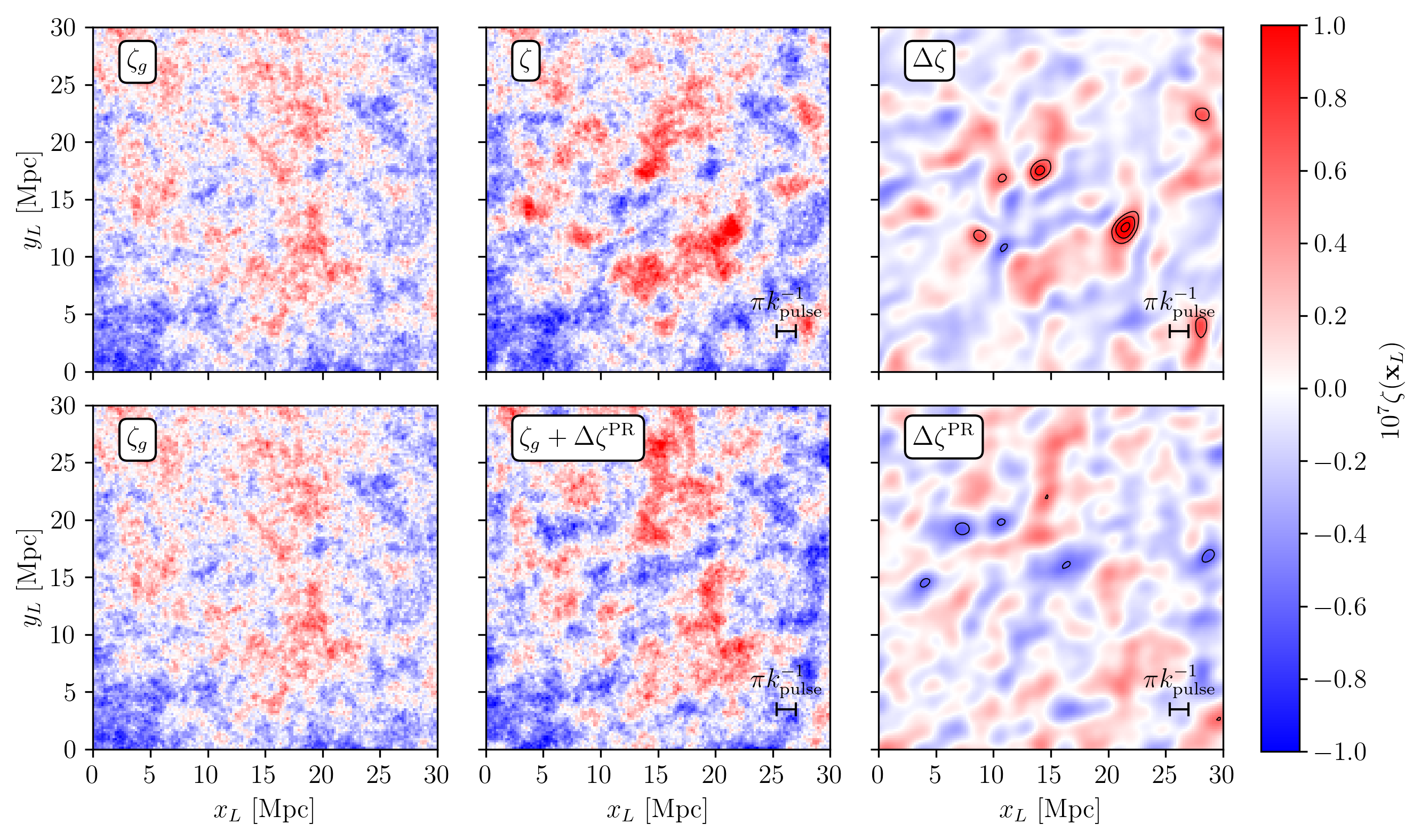}
    \vspace{-0.5cm}
    \caption{
        The \ping scalar field, $\zeta$, is contrasted with a phase-randomized Gaussian field. The top row from left to right shows the Gaussian field, $\zeta_g$, the \ping field, $\zeta$, and the response to the inflationary \ping potential feature, $\Delta\zeta$, with $\mathcal{D}_0 = 0.6$ and $\kpulse^{-1} = 5\times10^{-1/2} ~ \mathrm{Mpc}$. The half-wavelengths, $\pi\kpulse^{-1}$, are shown as this corresponds roughly to the diameter of a \ping peak. The bottom row from right to left shows phase-randomized perturbation, $\Delta\zeta^\mathrm{PR}$, the phase-randomized field added to the identical Gaussian field, $\zeta_g$, and the Gaussian field again. The black contours in $\Delta\zeta$ and $\Delta\zeta^\mathrm{PR}$ show $\pm2\sigma_{\zeta_g}, \pm3\sigma_{\zeta_g}, \pm4\sigma_{\zeta_g}$. The prominent peaks in $\Delta\zeta$ and absent in $\Delta\zeta^\mathrm{PR}$ show how power spectrum-only analyses of \ping features cannot capture their complete \nong effects.
    }
    \label{fig:phase scrambled fields}
\end{figure}

\begin{figure}
    \centering
    \vspace{-1em}
    \includegraphics[width=0.67\linewidth]{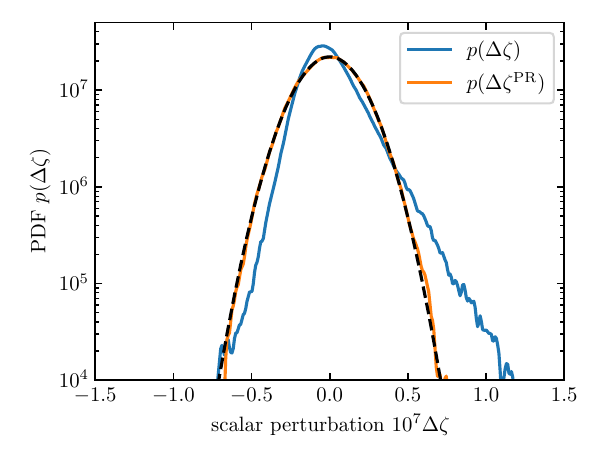}
    \vspace{-0.5cm}
    \caption{
        The probability density functions are compared for the \ping response, $\Delta\zeta$, and the phase-randomized field, $\Delta\zeta^\mathrm{PR}$, as shown in Figure \ref{fig:phase scrambled fields}. Each \PDF is measured from a $532^2$-voxel realization of the respective field. The dashed line is the best-fit Gaussian to the probability density function, $p(\Delta\zeta^\mathrm{PR})$. The phase-randomized field is Gaussian distributed, whereas the \ping field is heavily skewed.
    }
    \label{fig:phase scramble histogram}
\end{figure}

To illustrate the difference between field-level \ping and a \GRF with the same power spectrum as the \ping field, consider that the Fourier transform of the \ping perturbation field can be decomposed into a mode and the complex exponential of a phase, $\Delta\tilde{\zeta}(\mathbf{k}) = A_\mathbf{k} e^{i\theta_\mathbf{k}}$. The power spectrum, $\langle|\Delta\tilde{\zeta}^\mathrm{PR}|^2\rangle$, depends only on the modes because the randomly distributed phases will cancel when integrated over. If the phases are replaced by uniformly distributed random phases, the power spectrum is unaffected, but higher-order correlations are zeroed,
\begin{equation}
    \label{eq:phase scramble}
    \Delta\tilde{\zeta}(\mathbf{k}) ~ \longrightarrow ~ \Delta\tilde{\zeta}^\mathrm{PR}(\mathbf{k}) \equiv | \Delta\tilde{\zeta}(\mathbf{k})| e^{i\theta_\mathbf{k}^\mathrm{PR}}, \quad \theta_\mathbf{k}^\mathrm{PR} \sim U[0,2\pi), \quad \theta_\mathbf{-k}^\mathrm{PR}=(\theta_\mathbf{k}^\mathrm{PR})^*,
\end{equation}
where the final condition enforces Hermitian symmetry to ensure that the inverse Fourier transform remains real-valued, $\mathcal{F}^{-1}[\Delta\tilde{\zeta}^\mathrm{PR}(\mathbf{k})](\mathbf{x}) \in \mathbb{R}$.

Figure \ref{fig:phase scrambled fields} shows the dominant Gaussian field, $\zeta_g$, the fully non-Gaussian field-level \ping field, $\zeta$, and the \ping response, $\Delta\zeta$, in the top row. The bottom row shows the phase-randomized \ping response $\Delta\zeta^\mathrm{PR}$, and total phase-randomized total field $\zeta_g+\Delta\zeta^\mathrm{PR}$, as well as the Gaussian, $\zeta_g$, which is repeated for ease of visual comparison. Each of these fields has zero mean, and in a heat map of any mean-zero \GRF, $f$, there must be an equal number of pixels with $f=+f_0$ and $f=-f_0$. Thus, a non-Gaussian field, $\Delta\zeta$, can be recognized by a significant asymmetry in total volume of regions with an overdensity $\Delta\zeta_0$ compared to the volume with underdensity $-\Delta\zeta_0$. To highlight this asymmetry, extreme peaks in $\Delta\zeta$ and $\Delta\zeta^\mathrm{PR}$ are highlighted with black contour lines marking $\pm2\langle\zeta_g^2\rangle^{1/2}, \pm3\langle\zeta_g^2\rangle^{1/2},\pm4\langle\zeta_g^2\rangle^{1/2}$. While the phase-randomized field has a few peaks reaching $\pm2\sigma_{\zeta_g}$, the \ping field has multiple very prominent peaks. Thus, it is immediately clear by visual inspection that the $\Delta\zeta$ field is non-Gaussian. This is confirmed in Figure \ref{fig:phase scramble histogram} which contrasts the \PDF of $\Delta\zeta$ and $\Delta\zeta^\mathrm{PR}$. The latter closely matches a Gaussian, the former has a significant tail at high $\Delta\zeta$ which accounts for the ultra-prominent peaks observed in Figure \ref{fig:phase scrambled fields}. Note that we have chosen an extreme \ping case here to increase visibility of the \nong in teh fields. The \ping parameter space allows for much subtler effects that would not be so visibly striking.

By construction, the power spectra are equal for the perturbation fields, $\langle|\Delta\tilde\zeta|^2\rangle(k)=\langle|\Delta\tilde\zeta^\mathrm{PR}|^2\rangle(k)$. As illustrated in Figure \ref{fig:PING vs CMB matter power}, \ping power spectra have a spike-like feature in the power spectrum that peaks at a wavenumber $\kpulse$. Therefore, excess structure is expected to form with a wavelength $\lambda = 2\pi\kpulse^{-1}$. The half wavelength can be seen as a feature scale, $R_\mathrm{pulse}=\pi\kpulse^{-1}$, as it represents the distance from peak to trough, or the diameter of a \ping peak measured from node to node. Here, with a peak at $\kpulse^{-1}=5\times10^{-1/2}~\mathrm{Mpc}$, both fields in Figure \ref{fig:phase scrambled fields} have a characteristic scale of $\pi\kpulse^{-1}\sim10~\mathrm{Mpc}$. Any much larger or much smaller fluctuations are suppressed in $\Delta\zeta$ because they are subtracted out with $\zeta_g$, so visually, both the \ping response and phase-randomized fields have a similar ``smoothness''. The \nong field, $\Delta\zeta$, however may have higher-order connected correlations (bispectrum, trispectrum, and so on), so prominent peaks may be seen to cluster together more. In contrast, the Gaussian, $\Delta\zeta^\mathrm{PR}$, is visually more homogeneous. The higher-order correlators will therefore not necessarily be the most sensitive probes of \ping. Note that an extreme case of the \ping feature amplitude, $\mathcal{D}_0$, was chosen so that the \ping field will be visibly distinguishable from the Gaussian and phase-randomized fields.

We chose to phase-randomize the perturbation field, $\Delta\zeta$, so that it is easier to distinguish the contributions from $\Delta\zeta$ and $\Delta\zeta^\mathrm{PR}$ when added to $\zeta_g$. As a result, the cross-power of $\zeta$ and $\zeta_g+\Delta\zeta^\mathrm{PR}$ are not by definition zero. However, because $\Delta\zeta$ is sourced by a field that is independent of $\zeta_g$, cross-correlations should be minimal. These figures are used for the purpose of illustration, for a statistical study, the total fields should be phase-randomized.

The differences between $\Delta\zeta$ and $\Delta\zeta^\mathrm{PR}$ demonstrate that the \MCMC power spectrum analysis of Figure \ref{fig:CobayaMCMC} does not capture the full \ping effect. We thus turn to the \pkp simulations, which can realize cosmologies with field-level \png.

\subsection{Effects of field-level \ping in the \pkp simulations}\label{subsec:field level ping in pkp}

\begin{figure}
    \centering
    \vspace{-1em}
    \includegraphics[width=0.44076\linewidth]{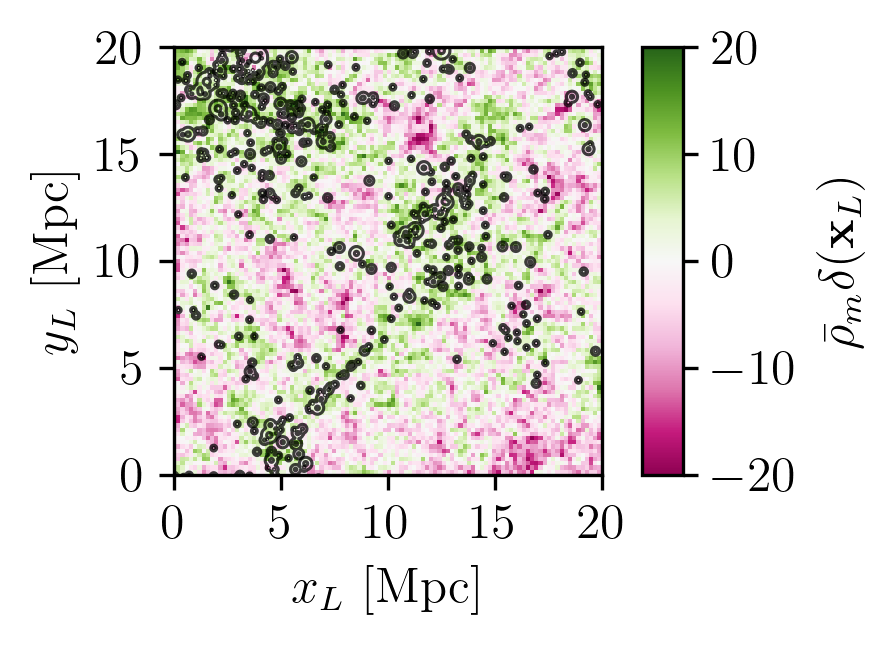}
    \vspace{-0.25cm}
    \caption{
        Heat map of linear matter field for a \pkp run with Gaussian initial conditions. \pkp halos are overlaid in black contour lines denoting mean densities $10^7$, $10^{12}$ and $3.16\times10^{13} ~ M_\odot\mathrm{Mpc}^{-3}$, averaged over a slab of thickness 10 Mpc. We emphasize that this is the $z=0$, Lagrangian positions, $\mathbf{x}_L$, such that halos are not displaced to their final positions in order to better overlap with the underlying matter field.
    }
    \label{fig:gaussian fields}
\end{figure}
\begin{figure*}
    \centering
    \vspace{-1em}
    \includegraphics[width=1\linewidth]{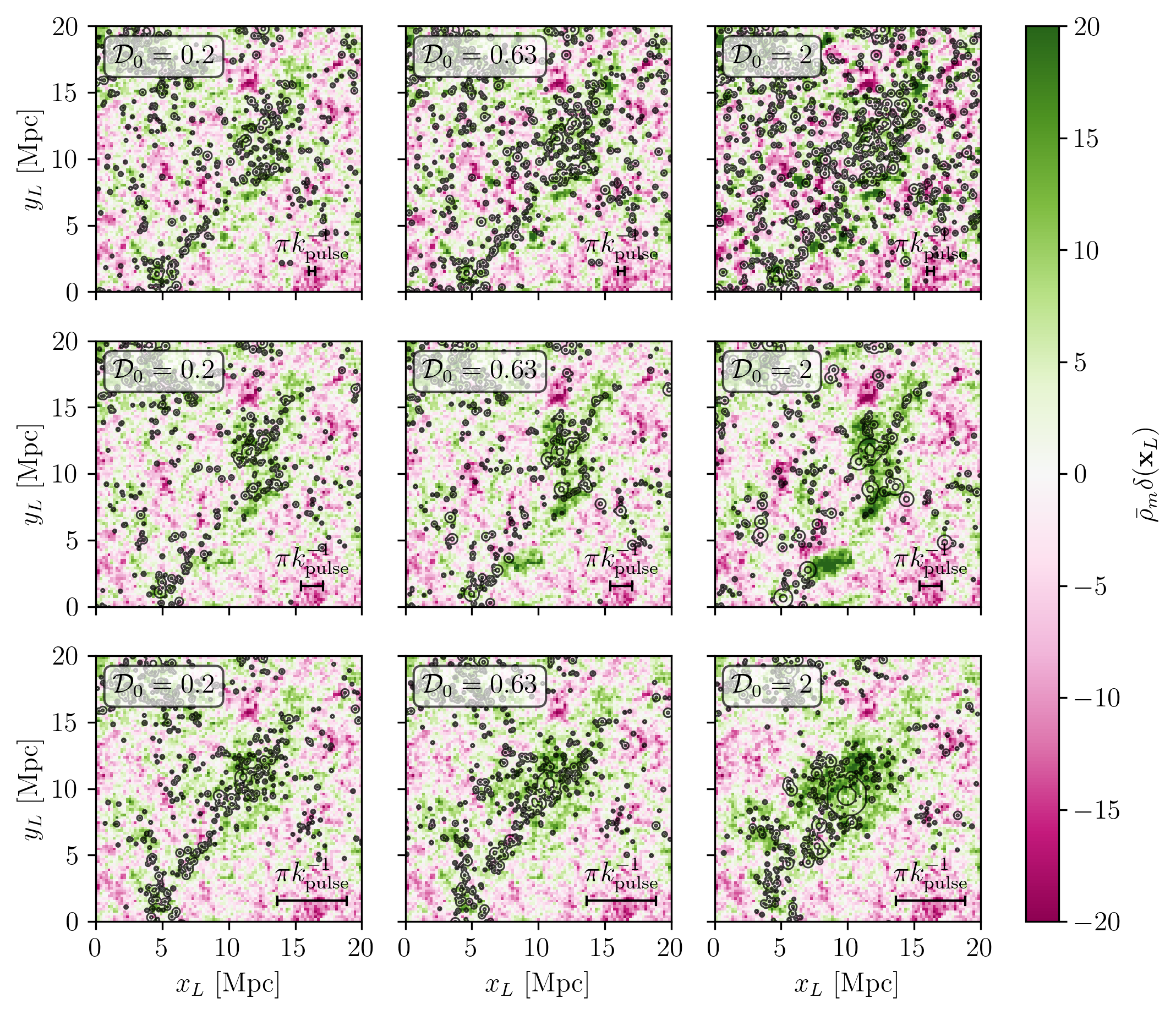}
    \vspace{-0.5cm}
    \caption{Heat map of linear matter field for \pkp run with \ping initial conditions. \pkp halos are overlaid in black contour lines denoting mean densities $10^7$, $10^{12}$ and $3.16\times10^{13} ~ M_\odot\mathrm{Mpc}^{-3}$, averaged over a slab of thickness 10 Mpc. Note that the volume corresponds to that of the primordially Gaussian \pkp run shown in \ref{fig:gaussian fields}. Rows from top to bottom are \ping matter fields with $\kpulse$ values of $6$, $6\times10^{-0.5}$ and $6\times10^{-1} ~ \mathrm{Mpc}^{-1}$. Columns from left to right have $\mathcal{D}_0$ values $2\times10^{-1}$, $2\times10^{-0.5}$ and $2$. The half-wavelength, $\pi\kpulse^{-1}$, is shown in each to indicate the characteristic scale of the \ping. Overdensities of size roughly $\pi\kpulse^{-1}$ are observed particularly prominently as $\mathcal{D}_0$ is increased. The halos trace the overdensities of the matter field for larger scale \ping. This trend is not as visually obvious for smaller scale \ping due to the halo slab thickness. Figure \ref{fig:ping fields zoom in} shows a thinner slice of the smallest scale \ping for reference. Excess halos with size of order $\pi\kpulse^{-1}$ are observed in each case.}
    \label{fig:ping fields}
\end{figure*}
\begin{figure*}
    \centering
    \vspace{-0.5em}
    \includegraphics[width=1\linewidth]{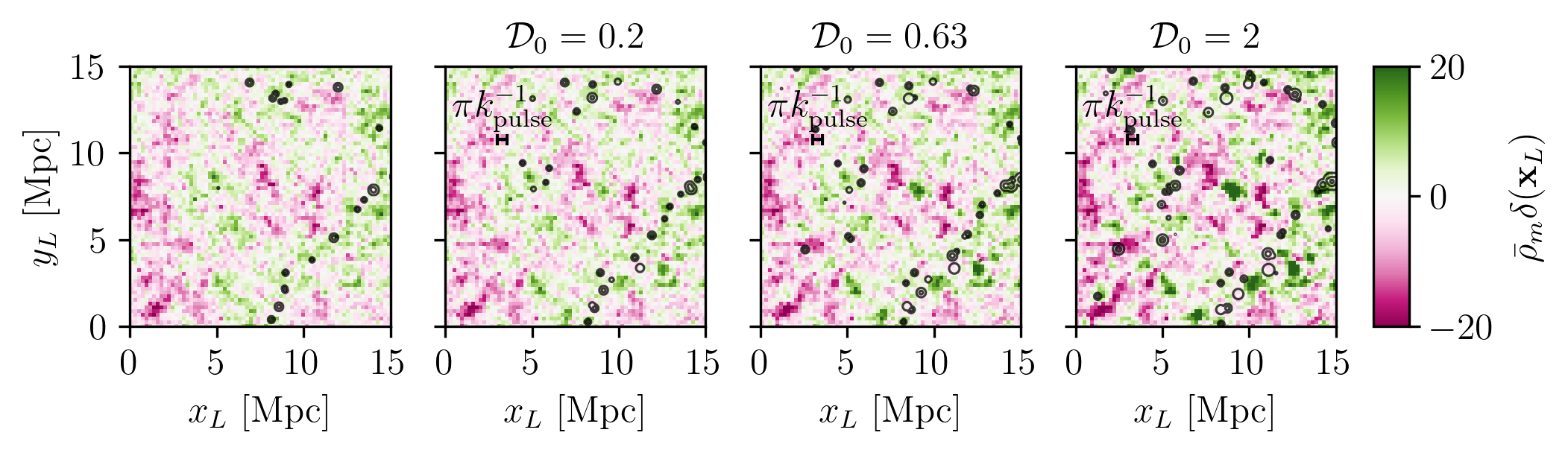}
    \vspace{-0.5cm}
    \caption{
        Heat map of linear matter field comparing Gaussian initial conditions and \ping with $\kpulse=6~\mathrm{Mpc}^{-1}$. Halo field overlaid in black contour lines denote mean halo densities of $1\times10^7$ and $1\times10^{12} ~ M_\odot\mathrm{Mpc}^{-3}$, averaged over a slab of thickness 1 Mpc. This thinner slab depth than was shown in Figures \ref{fig:gaussian fields} and \ref{fig:ping fields} is chosen to more closely match the \ping scale, $\pi\kpulse^{-1} \sim 0.52~\mathrm{Mpc}$. Thus excess halos that appear in the \ping maps can be more precisely connected to \ping features. The left panel shows the Gaussian field and halos, the remaining panels from left to right have $\mathcal{D}_0$ values $2\times10^{-1}$, $2\times10^{-0.5}$ and $2$. Excess halos are seen to more closely trace out overdense regions, though the largest \ping overdensity peaks in the matter field do not necessarily align with the halos.
    }
    \label{fig:ping fields zoom in}
\end{figure*}

The \pkp simulations produce a catalogue of \DM halos from a given linear matter field. Response functions can be applied to halo catalogues to realize mock sky maps as we show in Section \ref{sec:cii mocks}, but one can also study the statistics of the halos alone.

For comparison with the \pkp initial conditions matter field, \DM halo catalogues can be transformed into a field. This is achieved by 1) binning halos by the top-hat smoothing radius at which they were identified in the \pkp calculation, $R_{\mathrm{halo,TH}}$, then 2) for each bin, populating a 3D field with point sources of mass $M_\mathrm{halo} = \bar{\rho}_{m,0} \frac{4}{3} \pi R_\mathrm{halo,TH}^3$, and 3) smoothing each 3D field at the bin scale, $R_\mathrm{halo,TH}$, and finally 4) adding together the fields for each bin. For use as a visual aid and to study differences between halo fields, it is sufficient to simply use a Gaussian smoother, but for a more physical treatment, one could implement a more physically motivated halo profile, such as Einasto or NFW \citep{einasto_1965,NFW_1997}. For Gaussian smoothing, the filter scale is chosen to preserve the second moment obtained with top-hat smoothing, $R_\mathrm{halo,G} = R_\mathrm{halo,TH} / \sqrt{5}$. In Figure \ref{fig:gaussian fields}, this halo field is shown as black contour lines overlaying a $(20~\mathrm{Mpc})^2$ slice of the matter field, represented as a heat map. Halos from which we derive the density field, $\rho_\mathrm{halo}$, are shown in Lagrangian space, $\mathbf{x}_L$, so as to better align with the matter field, $\bar{\rho}_m\delta(\mathbf{x}_L)$. Halos can be seen to trace the overdensities of the matter field. Superclusters are seeded by large-scale regions of extreme matter overdensity, for example, at $(x_L,y_L)=(5~\mathrm{Mpc},17~\mathrm{Mpc})$. Conversely, large-scale underdensities in $\bar{\rho}_m\delta(\mathbf{x})$ seed few halos, leading to the formation of voids, for example at $(17~\mathrm{Mpc},0~\mathrm{Mpc})$.

Figure \ref{fig:ping fields} shows the same $(20~\mathrm{Mpc})^2$ slice of the matter and halo fields as shown in Figure \ref{fig:gaussian fields}, but includes full, field-level \ping in the initial matter field. The rows from top to bottom, respectively, have $\kpulse$ values of $(\frac{1}{6} ~ \mathrm{Mpc})^{-1}$, $(\frac{1}{6}\times10^{1/2} ~ \mathrm{Mpc})^{-1}$ and $(\frac{1}{6}\times10 ~ \mathrm{Mpc})^{-1}$. The columns from left to right have \ping with amplitude, $\mathcal{D}_0$, values of $2\times10^{-1}$, $2\times10^{-1/2}$ and $2$. For the large-scale \ping effect ($\pi\kpulse^{-1} \simeq 5.24$ bottom row), increasing $\mathcal{D}_0$ leads to more clustering, and therefore a small number of exceptionally large halos form with sizes comparable to the \ping scale, $R_\mathrm{halo,TH} \sim \pi\kpulse^{-1}$.

Each slice of the halo catalogue in Figure \ref{fig:ping fields} is a projection of a slab with a thickness of 10 Mpc. For smaller scale \ping shown in the top row, the \ping scale, $\pi\kpulse^{-1}\simeq0.524~\mathrm{Mpc}$ is much smaller than the slab thickenss, so the excess halos do not seem to trace out structure as well, instead just appearing as nearly uniform. To alleviate this confusion, in Figure \ref {fig:ping fields zoom in}, we overlay a thinner 1-Mpc slab of the halo catalogue on a smaller region of the matter field. In this figure, it becomes more clear that excess \ping halos are forming in overdense regions of the matter field, however they do not perfectly align, possibly due to tidal effects of these very small \ping peaks.

\begin{figure}
    \centering
    \vspace{-1em}
    \includegraphics[width=0.75\linewidth]{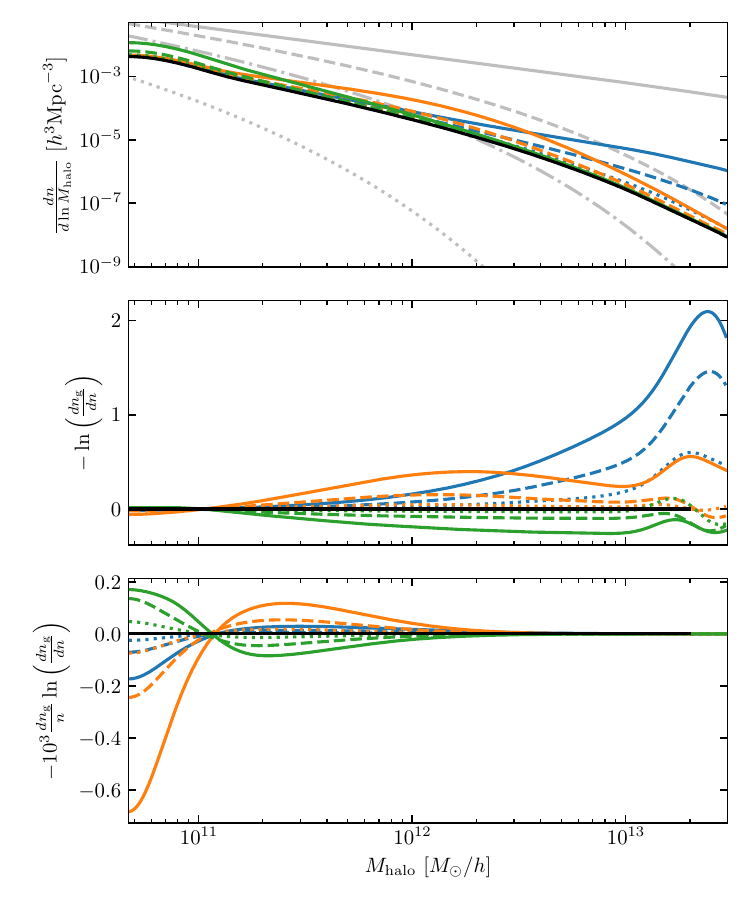}
    \vspace{-0.25cm}
    \caption{
        Halo mass functions and their relative entropy. The black line is the fiducial, primordially Gaussian result. The blue, orange and green curves represent \ping results with $\kpulse$ of $6\times10^{-1}$, $6\times10^{-1/2}$ and $6$ $\mathrm{Mpc}^{-1}$. The dotted, dashed and solid lines have $\mathcal{D}_0$ of $2\times10^{-1}$, $2\times10^{-1/2}$ and $2$. As suggested visually in Figures \ref{fig:ping fields} and \ref{fig:ping fields zoom in}, excess halos form with masses associated with the \ping scale in abundances positively correlated with the \ping amplitude $\mathcal{D}_0$. The relative entropy measure, $-\ln(dn_g/dn)$, emphasizes the surprise relative to the expected number of  high mass halos, and the measure, $-10^3(dn_g/dn)\ln(dn_g/dn)$, emphasizes the surprise in low mass halos. 
    }
    \label{fig:hmf}
\end{figure}

Halo catalogues can be compared statistically between realizations with the \HMF. The top panel in Figure \ref{fig:hmf} shows one form of the \HMF that describes how the number density of halos of mass $M_\mathrm{halo}$ changes with the natural logarithm of the halo mass. The solid black curve shows the \HMF for a \pkp simulation with purely Gaussian initial conditions. The green, orange and blue curves show large to small scale \ping features, respectively with $\kpulse$ values of $(\frac{1}{6} ~ \mathrm{Mpc})^{-1}$, $(\frac{1}{6}\times10^{1/2} ~ \mathrm{Mpc})^{-1}$ and $(\frac{1}{6}\times10 ~ \mathrm{Mpc})^{-1}$), and the solid, dashed and dotted coloured curves are high to low \ping feature amplitude, respectively with $\mathcal{D}_0$ of $2\times10^{-1}$, $2\times10^{-1/2}$ and $2$. As one might expect after examining the fields, \ping with a smaller scale power spectrum feature tends to produce an excess of smaller mass halos, and a larger feature leads to an excess of larger halos. Additionally, as the strength of the feature increases, more excess halos are formed.

It should be noted that these \HMF{}s are made from \pkp light-cone halo catalogues, meaning that all of the \DM halos in the catalogue are not at the same redshift. This is a more natural method of accounting for statistical halos masses as it ties more closely to observable tracers of \DM halos such as \LRGs. We leave direct comparisson of \pkp halos to \LRG catalogues such as those from \cite{DESI_2025a} to future work. Theory \HMF{}s typically assume all halos are at a single redshift. We show single-redshift \HMF{}s for contrast as well in the top panel in Figure \ref{fig:hmf} using the \texttt{hmf} code \citep{murray_2013} with $dn(M,z)$ from \cite{behroozi_2013} at $z$ of 0, 3, 5.5, and 8 (solid, dashed, dash-dotted and dotted grey curves respectively). Because there are a larger number of halos at the low mass end, the smallest halos are ubiquitous in a multi-Gpc volume. The number of halos of any given mass also decreases as redshift increases, since there has been less time for halos to form at higher redshift. In a $3.5 \le z \le 8$ light-cone \HMF, we therefore expect the high-mass end to be similar to the $z=8$ \HMF because there are fewer halos of this size. At the low mass end, because the small halos are ubiquitous, the light-cone \HMF should deviate toward the $z=3.5$ \HMF. This is precisely what is observed in the Gaussian light-cone \HMF.

Note the subtle bump in the Gaussian light-cone \HMF at $z\lesssim10^{11}$, which is due to resolution effects that arise during the merging and exclusion phase of the \pkp calculation. This could in principle be corrected with abundance matching techniques, but for the purposes of this work we assume the effect should be minimal as the \cii signal is dominated by emission from high-mass halos.

The second and third panels of Figure \ref{fig:hmf} show measures of the deviation from the Gaussian case. The form of these measures is chosen in analogy to surprisal and relative entropy, which are discussed further in Section \ref{subsec:Srel}. Note that the first measure, $-\ln(dn_g/dn)$, emphasizes deviations from the fiducial Gaussian case at high halo mass, and the second, $-dn_g\ln(dn_g/dn)/n$, emphasizes the low halo mass deviations.

\begin{figure}
    \centering
    \vspace{-0.5em}
    \includegraphics[width=1.0\linewidth]{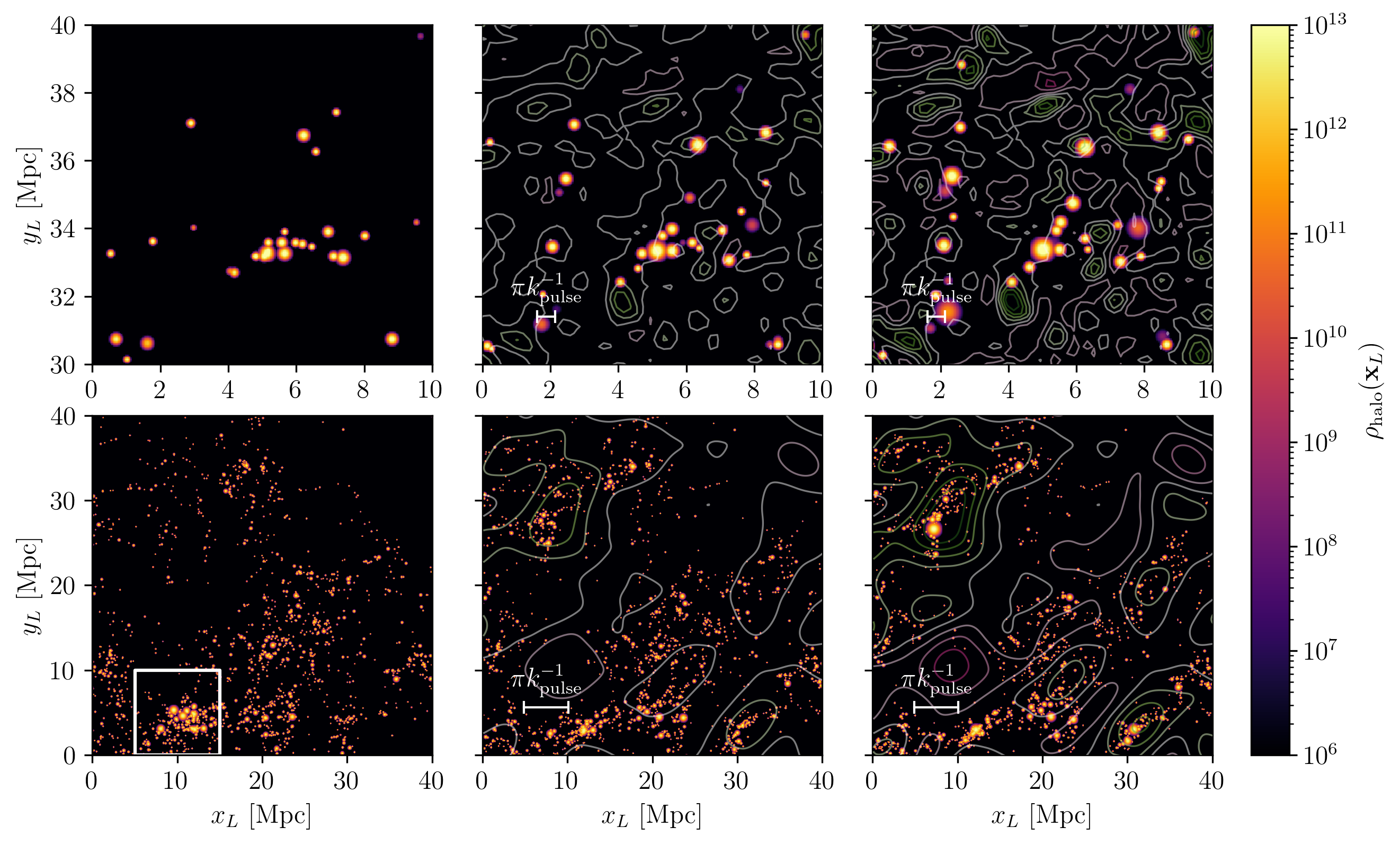}
    \vspace{-0.5cm}
    \caption{
        The effect of \ping on the \DM halo catalogue. The left column corresponds to a primordially Gaussian halo catalogue, the centre column a catalogue with moderate \ping, $\mathcal{D}_0=2\times10^{-1/2}$, and the right column a catalogue with strong \ping, $\mathcal{D}_0=2$. The bottom row is a $40 \times 40 \times 10 ~ \mathrm{Mpc}^3$ volume of a halo catalogue and the top row is a thinner $10\times10\times1 ~ \mathrm{Mpc}^3$ volume centred on the white square in the bottom left panel. The \ping catalogues in the top row has small-scale features, $\kpulse^{-1} = 0.16\bar{6} ~ \mathrm{Mpc}$, whereas the bottom row has large-scale features, $\kpulse^{-1}=1.6\bar{6} ~ \mathrm{Mpc}$. Contours represent the \ping response fields, $\Delta\zeta$. These use a common colour map with 8 evenly spaced contours from $-1.5 \sigma_{\zeta_g}$ in magenta to $2 \sigma_{\zeta_g}$ in dark green. In the large-scale \ping case, excess clusters can form in a void region if a strong \ping feature occurs, such as at $(10~\mathrm{Mpc},27~\mathrm{Mpc})$. In the small-scale \ping case, the effect is subtler, with less prominent $\Delta\zeta$ peaks pushing already mildly overdense $\zeta$ regions through the threshold to form a halo, so halos do not correlate as strongly with $\Delta\zeta$. An animated version of this plot is available at \href{https://uoft.me/webskycii}{https://uoft.me/webskycii}.
    }
    \label{fig:PING in void}
\end{figure}

\ping produces isolated peaks in the $\zeta$ field that are uncorrelated with the Gaussian field, $\zeta_g$. This has a number of potentially measurable consequences in the spatial distribution of \DM halos that would not be seen in the \HMF. Because the response, $\Delta\zeta$, is uncorrelated with $\zeta_g$, \ping can produce isolated sharp $\zeta$ peaks in low-$\zeta_g$ regions. If the \ping scale, $R_\mathrm{\ping}\sim\pi\kpulse^{-1}$, is smaller than the void, then for sufficiently strong $\mathcal{D}_0$, this could produce excess void halos. 

Figure \ref{fig:PING in void} shows the effect of \ping with $\pi\kpulse^{-1}] = 5.24 ~ \mathrm{Mpc}^{-1}$ and $0.524 ~ \mathrm{Mpc}^{-1}$ and varying $\mathcal{D}_0$ on a $40 \times 40 \times 10 ~ \mathrm{Mpc}^3$ slice of the halo field at redshift $z\sim3.5$. The halo fields, shown as heat maps, are made in the same fashion as Figures \ref{fig:gaussian fields} and \ref{fig:ping fields}. The left-most panels show the primordially Gaussian case, while the middle and right panels have respective \ping with $\mathcal{D}_0=2\times10^{-1/2}$ and $2$. The $\Delta\zeta$ field is overlaid with the 8 contours evenly spaced in $\Delta\zeta$ from $-1.5$ to $2$ times the standard deviation of the $\zeta_g$ field. The segment of the \DM halo catalogue in the bottom row was chosen because upper left and right corners show significant voids of volume around $(20~\mathrm{Mpc})^3$ and at comoving position $(x,y) \sim (10~\mathrm{Mpc},0.3~\mathrm{Mpc})$ there is a supercluster. The $\Delta\zeta$ field has a $>2\sigma_{\zeta_g}$ prominence at $(x,y)\sim(8~\mathrm{Mpc},28~\mathrm{Mpc})$ near the centre of a void. As hypothesized, this leads to a cluster of void halos growing in mass with $\mathcal{D}_0$. The top row shows a thinner 1-Mpc thick slice of the halo catalogue centred on the white square shown in the bottom left panel. This region was chosen simply as it has more halos making the effects of adding in \ping more statistically significant in this visual representation. Some excess halos appear where there are peaks in $\Delta\zeta$, some also appear in the vicinity of peaks, suggesting that underlying overdensities in $\zeta_g$ were given just enough excess that the nearby \ping peak pushed them over the threshold to form halos.

The $\Delta\zeta$ peak also has an attractor effect, drawing in existing halos or absorbing them into the \ping halos. Another interesting effect arises due to the significant $\Delta\zeta$ underdensity at $(x,y) \sim (10~\mathrm{Mpc},10~\mathrm{Mpc}$. Quite the opposite of the $\Delta\zeta$ peaks, this feature acts as a repulsor, repelling existing halos and compressing the supercluster at $(x,y) \sim (10~\mathrm{Mpc},0.3~\mathrm{Mpc})$ into a more wall-like structure.

Like clusters, voids do not have well-defined boundaries, and as a result void halos are also not well defined. Voids could be selected as regions of mass density less than the mean density minus a chosen multiple of $\sigma$ by smoothing the halo field on a very large scale, $>10~\mathrm{Mpc}$. Then, one could measure the \HMF or two-point correlation functions in the void regions and compare between simulations. The halo-halo correlation function could also be useful in studying void halos and other clustering effects such as attractors and repulsors. These measures would be more useful with direct comparison to a galaxy catalogue, so we leave it to future work focused on cross-correlations with galaxies.

\section{Constraints on \texorpdfstring{\LIM}{LIM} models and \texorpdfstring{\ping}{PING} from \texorpdfstring{\cii}{[CII]} mocks}\label{sec:cii mocks}

\begin{figure*}
    \centering
    \vspace{-1em}
    \includegraphics[width=1\linewidth]{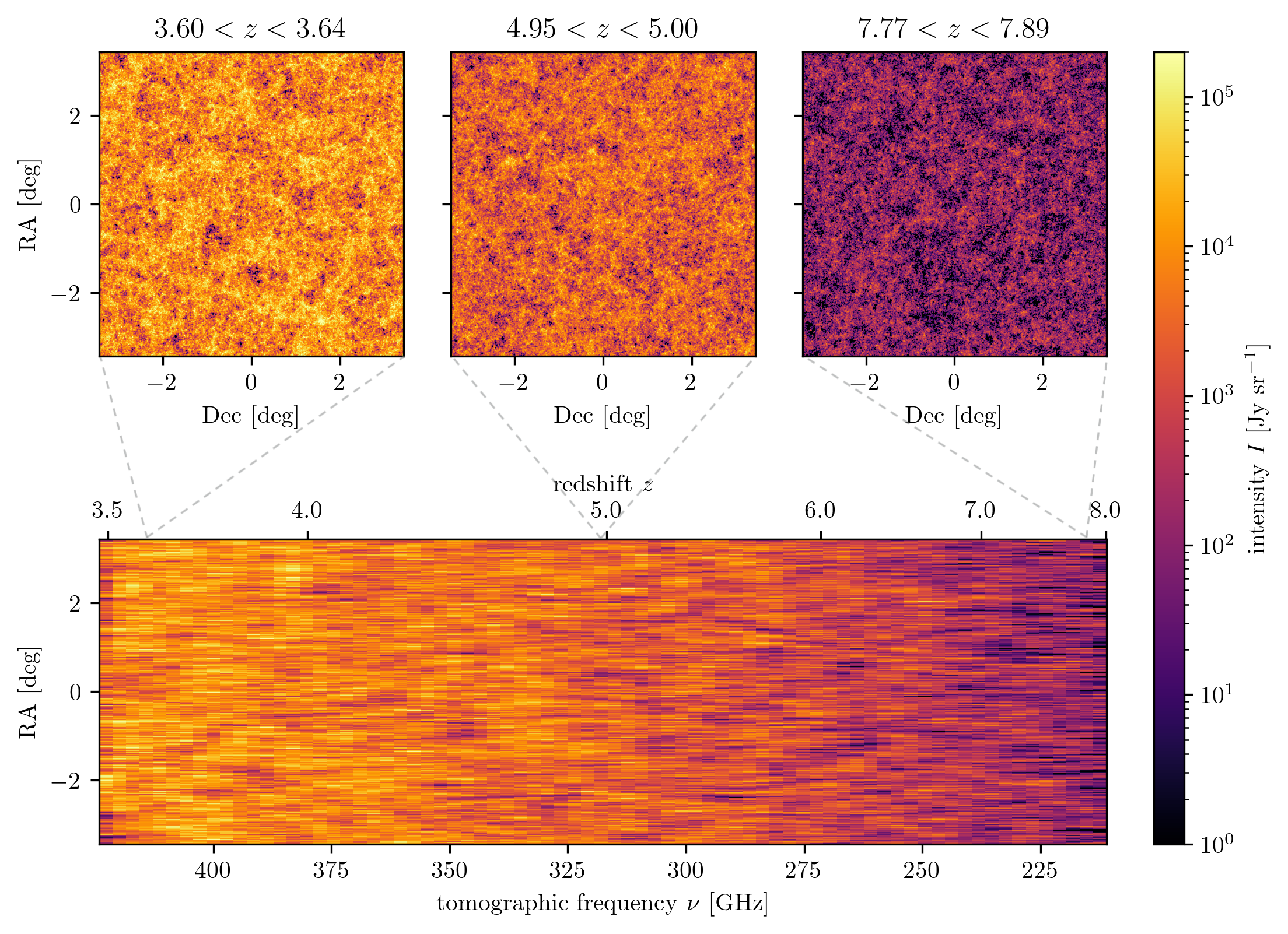}
    \vspace{-0.5cm}
    \caption{
        Tomographic \cii mock maps smoothed with a Gaussian filter at the CCAT telescope beam width of $FWHM=48''$. The top three panels show individual maps at specified three frequency bins corresponding to the redshifts shown above each. The bottom panel shows a strip from the centre of each frequency bin. The redshift eveolution of the cosmological \LIM signal is apparent in the gradual drop in mean intensity with increasing redshift. The correlation between the cosmological \LIM signal and the Cosmic Web is evident in the clustering observed in each of the highlighted redshift slices in the top row.
    }
    \label{fig:CII Mock Maps}
\end{figure*}

We apply the \ws \cii response functions described in Section \ref{subsec:CII I model} to a \pkp \DM halo catalogue to produce tomographic mocks. Figure \ref{fig:CII Mock Maps} shows an example of a tomographic \cii intensity cube. These mock maps are created from a \pkp \DM halo catalogue generated from a primordially Gaussian matter field and with the fiducial values of the \cii model parameters, $\alphacii$ and $\sigma_Z$, shown in the second column of Table \ref{tab:LIM Srel params}. This can be contrasted with additional \pkp catalogues with different \LIM parameters, \LambdaCDM model parameters, or with \png or \ping in the \pkp initial-conditions matter field, but to do so requires a summary statistic.

Because the cosmos around the time that the \CMB was released is relatively well-approximated by a \GRF, the power spectrum is a natural summary statistic. \LIM signals, however, originate from within highly non-Gaussian DM halos, so teh power spectrum is not an intuitively useful way of categorizing them. Further complications arise because the comoving volume of the voxels in a tomographic \LIM mock vary with redshift over the light cone. Prior work \cite{breysse_2017} shows that another one-point statistic, the \VID, performs perhaps surprisingly well as a summary statistic. This has been explored in much more detail in
\cite{breysse_2019,ihle_2019,sato-polito_2022,breysse_2023,chung_2023} and \citeHCBL.

\subsection{Relative entropy of the \texorpdfstring{\textup{\cii}}{[CII]} voxel intensity distribution}\label{subsec:Srel}

To compare the observed and expected VID, \cite{lee_2024} show that for a randomly distributed variable, $I$, (which we choose suggestively as the relevant variable later on is an intensity), the relative information content or ``surprisal'', $S(I)$, of the fiducial ``true'' \PDF, $p(I)$, compared to a \PDF, $q(I)$,
\begin{equation}
    \label{eq: srel}
    \dot{s}_\mathrm{rel}(I)
    \equiv
    S_p(I) - S_q(I)
    =
    - \ln \frac{p(I)}{q(I)},
\end{equation}
portrays well the far-tail difference between $p(I)$ and $q(I)$ if the \PDF{}s are unimodal with decaying tails. Note that in this work, $\dot{(~~)}=\frac{d(~~)}{dI}$ unless otherwise stated. Weighting the relative surprisal by the true \PDF damps the extreme tails, emphasizing the difference in the vicinity of the mean, thus capturing the variance, skewness and kurtosis, which encode non-Gaussianity for weakly non-Gaussian distributions. This statistic is the \PRE,
\begin{equation}
    \label{eq: dSrel}
    \dot{S}_\mathrm{rel}(I)
    \equiv
    - p(I) \ln \frac{p(I)}{q(I)}.
\end{equation}
This is defined so that its integral is the negative \KLD \citep{kullback-leibler_1951}. This unconventional negative signature is chosen to make the connection between the \PRE from $q(I)$ to the ``true'' $p(I)$ and the Shannon information of $p(I)$ minus that for $q(I)$. For the purposes of this work, the probability density functions in question are related, each representing a \VID for a \LIM model with some fiducial set of model parameters, $\bar{\bm{\lambda}}$, as compared with another set of model parameters, $\bm{\lambda}$, and thus the \PRE is
\begin{equation}
    \dot{S}_\mathrm{rel}(I;\bm\lambda,\bar{\bm\lambda})
    \equiv
    - p(I;\bar{\bm\lambda}) \ln \frac{ p(I;\bar{\bm\lambda}) }{ p(I;\bm\lambda) }
    .
\end{equation}
For logarithmically distributed variables, such as intensity, it makes sense to instead take the derivative with respect to $\ln I$, giving the \LPRE, $S'_\mathrm{rel}\equiv\frac{dS_\mathrm{rel}}{d\ln I}$. To quantify the amount of surprise added by a change in one of the model parameters, $\lambda_i$, we can consider the derivative of \PRE with respect to $\lambda_i$,
\begin{equation}
    \partial_{\lambda_i} \dot{S}_\mathrm{rel}(I;\bar{\bm\lambda})
    \equiv
    p(I;\bar{\bm\lambda}) \left. \frac{\partial}{\partial\lambda_i} \ln p(I;\bm\lambda) \right|_{\bm\lambda=\bar{\bm\lambda}}
    ,
\end{equation}
or similarly using the \LPRE.
\begin{figure}
    \centering
    \vspace{-6.5em}
    \includegraphics[width=0.9\linewidth]{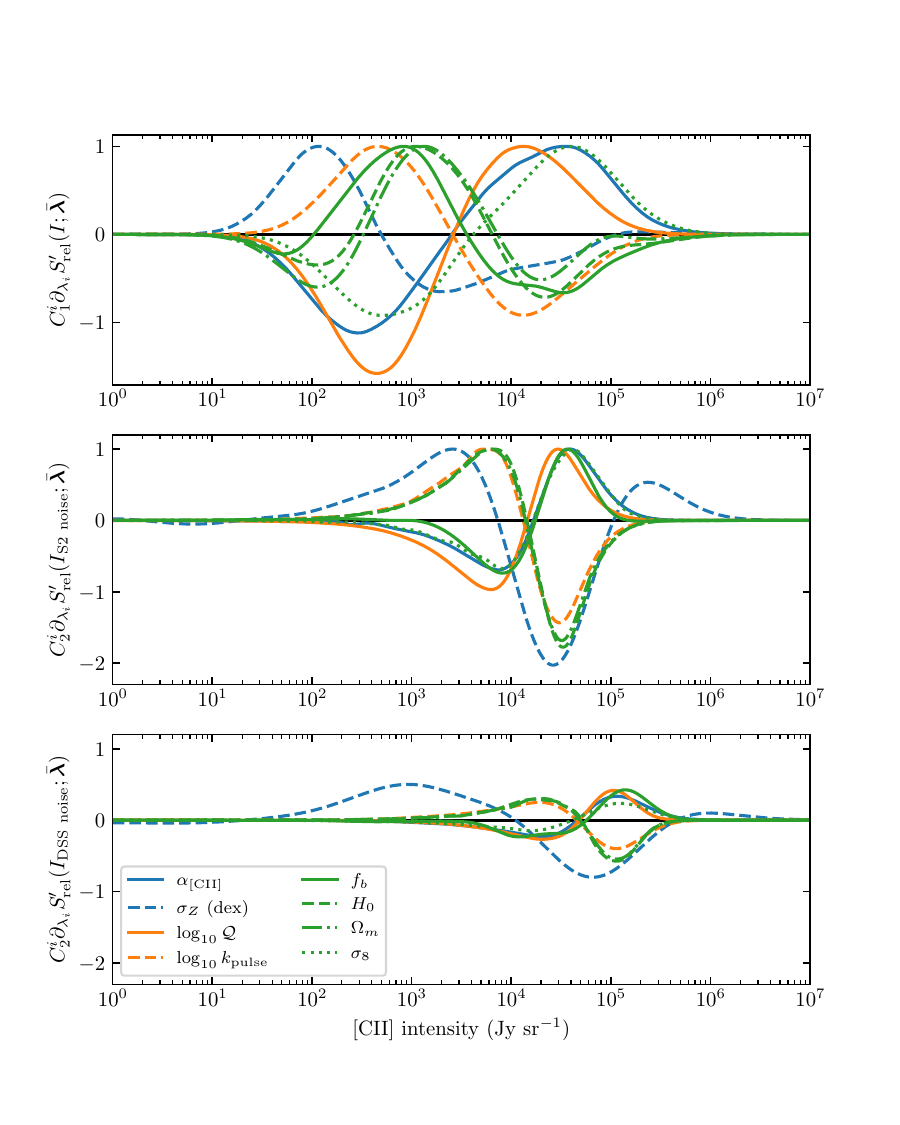}
    \vspace{-3em}
    \caption{
        The differential \LPRE with respect to \LIM, \LambdaCDM and \ping model parameters. Because model parameters span many orders of magnitude, absolute values of this information measure also span many orders of magnitude. We therefore include normalization factors $C^i$ for each model parameter, $\lambda_i$, which are listed in Table \ref{tab:LIM Srel params}. The top panel is theoretical noise-free differential \LPRE, the middle panel includes \Stwo noise and the bottom panel includes \DSS noise. Differential \LPRE encodes the information content of the signal as a function of intensity, thus distinct \LPRE curves will be easier to distinguish from one each other. The addition of noise significantly reduces the distinctiveness with \LPRE grouping into 3 or 4 similar shapes.
    }
    \label{fig:LIM Srel}
\end{figure}
Figure \ref{fig:LIM Srel} shows the change in information entropy with each of the \LambdaCDM, \LIM and \ping model parameters for a pure signal, and with \DSS and \Stwo noise models. Each differential \LPRE in the top panel depicting pure signal is multiplied by a factor, $C_1^i$, such that the maximum value is 1. This is done simply because the model parameters vary by several orders of magnitude. We do similarly in the second panel depicting the \Stwo noise-added signal with factors, $C_2^i$. In the bottom panel depicting the \DSS noise, we again use the factors, $C_2^i$, to show how the differential \LPRE changes from the \DSS case to the \Stwo case. The values of these coefficients are listed in Table \ref{tab:LIM Srel params}. Note that $C_1^i$ and $C_2^i$ generally differ by a factor of a few, suggesting that the orders of magnitude of the parameter fiducial values, $\bar{\lambda}_i$, are the major contributors to these factors.

The amount of information about each model parameter contained in the cosmological \cii intensity signal is encoded as a function of intensity in these differential \LPRE distributions. If a given model parameter is particularly tied to larger halos for instance, you might expect a more dramatic differential \LPRE in higher intensity bins. Furthermore, if the $\partial_{\lambda_i} S'_\mathrm{rel}$ curves for all model parameters, $\lambda_i$, have noticeably different shapes, this would indicate that if we made a measurement of the \cii \VID and it differed from fiducial \VID, we could likely distinguish which parameters differed from the fiducial values. It is therefore encouraging that Figure \ref{fig:LIM Srel} is so busy, particularly in the \Stwo scenario. The metallicity scatter, for instance, has a very unique differential \LPRE in all cases, so we would expect effects of misjudging its mean value to be easily distinguished from other modelling issues. Similarly, both \ping parameters shown in these plots are at least moderately discrepant with the other differential \LPRE curves, indicating that these surveys, provided sufficient resolution, may be able to detect \ping in observations of the cosmological \cii signal.

The fiducial values of each of the \LIM, \LambdaCDM and \ping{} model parameters are also shown in table \ref{tab:LIM Srel params}. The choice of fiducial parameters is discussed in more detail in the next section.

\begin{table*}
    \centering
    \caption{The parameters $\lambda_i$ varied in Fisher analysis of Section \ref{subsec:fisher} with fiducial values $\bar{\lambda}_i$ and priors $\mathcal{I}_{\mathrm{prior},ii}$. The columns $C^i_1$ and $C^i_2$ show the scalar coefficients used in Figure \ref{fig:LIM Srel}. Finally, $\sigma_\mathrm{DSS}$ and $\sigma_\mathrm{S2}$ are the $1\sigma$ forecasted uncertainties shown in figure \ref{fig:marginalized DSS and S2 forecasts} (blank rows are prior driven). Posteriors for $\kpulse$ and $H_0$ are shown in bold face for emphasis. $\kpulse$ is a largely unconstrained early universe parameter that is here resolved to within an order of magnitude, and the constraint on $H_0$ is competitive with other measures.}
    \begin{tabular}{c!{\vrule width 1pt}cc|cc|cc}
        $\lambda_i$                     & $\bar{\lambda}_i$ & $\mathcal{I}_{\mathrm{prior},ii}$ &  $C_1^i$            & $C_2^i$             & $\sigma_\mathrm{DSS}$ & $\sigma_\mathrm{S2}$ \\
        \Xhline{1pt}
        $\alphacii$                     & $0.024$           & $(5\times0.024)^{-2}$             & $3.48\times10^{-1}$ & $2.50\times10^{-1}$ & $0.106$               & $0.072$              \\
        $\sigma_Z ~ (\mathrm{dex})$     & $0.4$             & $(5\times0.2)^{-2}$               & $5.29$              & $1.84\times10^2$    & $0.69$                & $0.46$               \\
        \hline
        $\log_{10}\mathcal{Q}$          & $-13.625$         & $(-14.0)^{-2}$                    & $8.03$              & $5.01$              & $7.33$                & $2.89$               \\
        $\log_{10}\kpulse$              & $\log_{10}(6)$    & $0$                               & $9.08$              & $5.24$              & $\textbf{2.14}$       & $\textbf{1.01}$      \\
        $m_\lambda$                     & $50$              & $(50)^{-2}$                       & $3.86\times10^2$    & $6.51\times10^2$    & $-$                   & $-$                  \\
        $m_\chi$                        & $25$              & $(25)^{-2}$                       & $4.66\times10^2$    & $1.49\times10^3$    & $-$                   & $-$                  \\
        \hline
        $f_b$                           & $0.1571$          & $(100\times0.0045)^{-2}$          & $1.23\times10^1$    & $9.97$              & $-$                   & $-$                  \\
        $\Omega_m$                      & $0.3117$          & $(100\times0.0073)^{-2}$          & $6.80$              & $2.21$              & $-$                   & $-$                  \\
        $\frac{H_0}{\mathrm{km/s/Mpc}}$ & $67.36$           & $(10\times0.73)^{-2}$             & $5.99$              & $1.92$              & $\textbf{1.7}$        & $\textbf{0.60}$      \\
        $\sigma_8$                      & $0.8111$          & $(100\times0.0060)^{-2}$          & $8.28$              & $5.20$              & $-$                   & $-$                  \\
    \end{tabular}
    \label{tab:LIM Srel params}
\end{table*}

In interest of keeping feasible the required number of \pkp and \ws simulations, we limited the number of model parameters varied. For instance, the differential \LPRE for most \cii model parameters were seen in \citeHCBL to have very similar shapes to either $\alphacii$ or $\sigma_Z$, so we simply fixed all but these two parameters. For a more thorough study, one could allow all of the \cii model parameters to vary.

\subsection{Fisher analysis for \textup{\CCAT \DSS} and \Stwo \texorpdfstring{\textnormal{\cii}}{[CII]} \texorpdfstring{\LIM}{LIM} surveys}\label{subsec:fisher}

\begin{figure*}
    \centering
    \vspace{-1em}
    \includegraphics[width=1.0\linewidth]{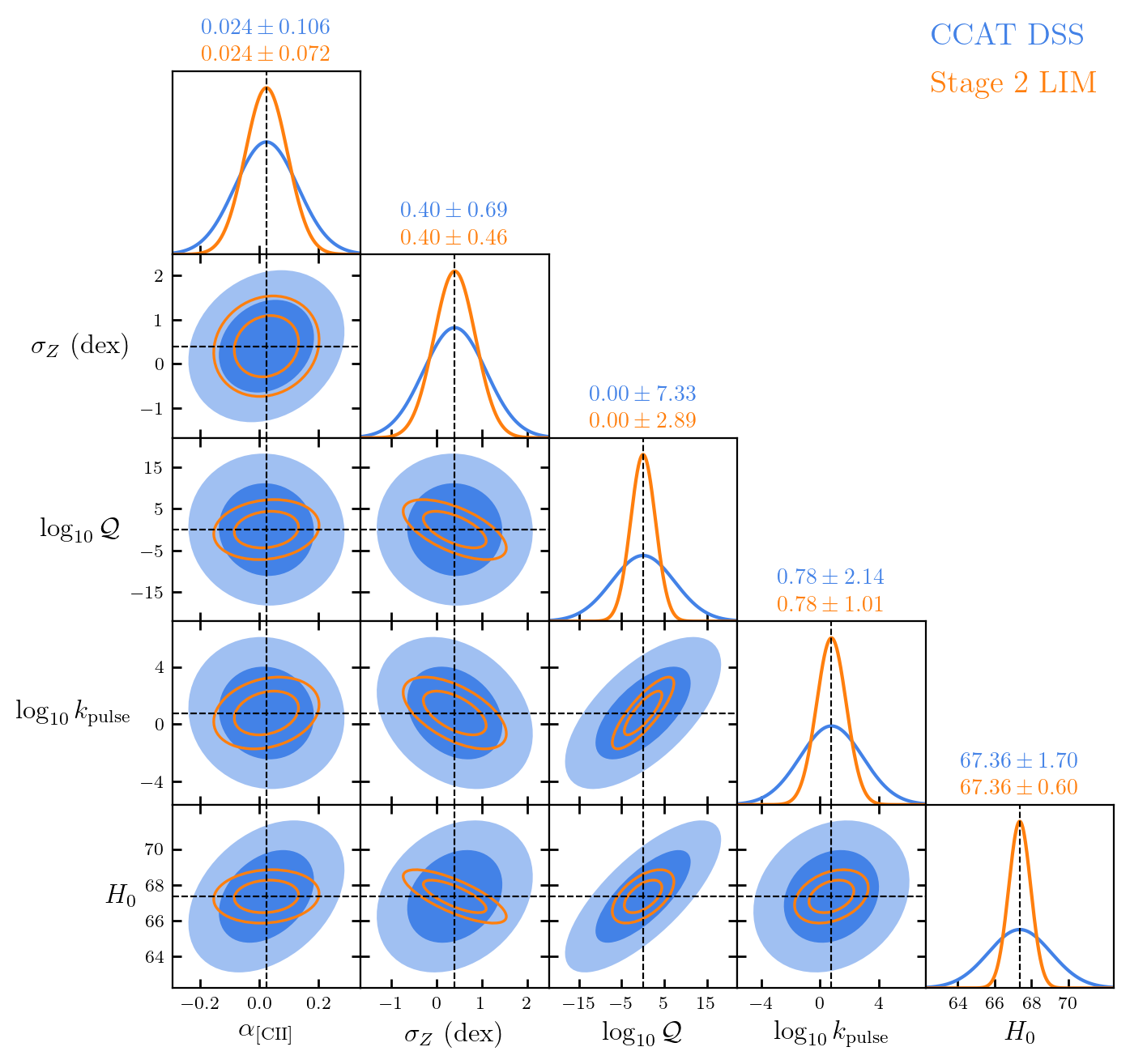}
    \vspace{-0.5cm}
    \caption{
        $1\sigma$ and $2\sigma$ confidence ellipses with \CCAT \DSS (blue solid ellipses) and \Stwo (orange contour lines) noise models. Additional parameters that were prior driven are marginalized over. The mean and $1\sigma$ uncertainties in each \PDF are shown above the diagonal panels. For the most part, contours are not strongly correlated or anti-correlated, which implies distinguishability in surveys. In particular, grouping into \LIM model parameters, ($\alphacii,\sigma_Z$), \ping model parameters, ($\log_{10}\mathcal{Q},\log_{10}\kpulse$) and the single remaining \LambdaCDM parameter, $H_0$, \DSS \LIM model parameters are not strongly correlated with \ping or \LambdaCDM parameters, meaning that novel effects in the cosmological \LIM signal will be distinguishable from new cosmology. \Stwo also has the potential to distinguish between \ping and variations in the \LambdaCDM model.
    }
    \label{fig:marginalized DSS and S2 forecasts}
\end{figure*}

To make forecasts for upcoming LIM surveys, we must quantify not just the amount of information the \VID, $p(I;\bm\lambda)$, holds about each parameter, $\lambda_i$, (which we have seen is encoded in the differential \LPRE of Figure \ref{fig:LIM Srel}), but also to what extent that information is independent for any two parameters, $\lambda_i$ and $\lambda_j$. This is achieved with the Fisher information matrix,
\begin{align*}
    \Fisherij{\bar{\bm{\lambda}}}
    &\equiv
    E\left[ \left( \frac{\partial \ln p(I;\bm{\lambda})}{\partial \lambda_i} \right) \left( \frac{\partial \ln p(I;\bm{\lambda})}{\partial \lambda_j} \right) \Bigg| \bar{\bm{\lambda}} \right]
    \\
    &=
    \int \frac{ \partial_{\lambda_i}\dot{S}_\mathrm{rel}(I;\bar{\bm{\lambda}}) \partial_{\lambda_j}\dot{S}_\mathrm{rel}(I;\bar{\bm{\lambda}})  }{ p(I;\bar{\bm{\lambda}}) } dI
    .
    \countme\label{eq:Fisher information matrix}
\end{align*}
The inverse of the Fisher information matrix places a lower bound on the covariance \citep{rao_1945,cramer_1946}. We can make a forecast for a survey using our model and priors for each of its parameters. The priors for our survey are constraints we already have on each of the model parameters. If the fiducial value of each of $N$ model parameters, $\bar\lambda_i$, has variance, $\sigma_i$, then the priors are a second Fisher matrix whose elements are the inverses of the prior covariances. In this work, priors are chosen based on a physics-informed estimate of the uncertainties from other experiments. If changing the value of a prior, $\mathcal{I}_{\mathrm{prior},ii}$, has a significant effect on the predicted uncertainty, $\sigma_i$, then the prediction is said to be ``prior driven''. This indicates that the survey does not have sufficient resolution to accurately constrain the corresponding model parameter, $\lambda_i$.

\citeHCBL showed that for most of the \cii model parameters, the relative entropy curves displayed a proportionality with either the overall \cii amplitude, $\alphacii$, or the matallicity scatter, $\sigma_Z$. For efficiency, we chose in this work to only consider $\alphacii$ and $\sigma_Z$ in our Fisher analysis, maintaining the fiducial values of $\alphacii=0.024$ and $\sigma_Z=0.4 ~ \mathrm{dex}$. To estimate an uncertainty, we consider that \cite{vizgan_2022} found $\log_{10}(L_\odot M_\mathrm{HI}/M_\odot \Lcii) = 1.7^{+0.4}_{-0.3}$, which corresponds to an uncertainty in $\alphacii$ of about $\sigma_{\alphacii}\simeq\pm0.018$. However, because we have chosen only to consider $\alphacii$ and $\sigma_Z$ in our Fisher analysis, we choose to take a loose prior, $\sigma_{\alphacii} = 5\times0.024$ to allow it to absorb some of the uncertainty on other model parameters. There is no uncertainty reported for the metallicity scatter, but given these works are done at $z=6$, whereas this work spans $z\in[3,8]$, we therefore choose a loose prior, $\sigma_{\sigma_Z}=5\times0.4$.

We consider a selection of cosmological parameters: the baryon fraction, $f_b = \Omega_b/\Omega_m$, the matter density fraction, $\Omega_m$, the Hubble constant, $H_0$, and the root-mean-square amplitude of the matter density fluctuations smoothed with a spherical top-hat filter of radius $8~h^{-1}\mathrm{Mpc}$, $\sigma_8$. The fiducial values of these are shown in Table \ref{tab:LIM Srel params}. For priors, we once again draw on the \planck Collaboration's 2018 results for TT,TE,EE+lowE+lensing, using the uncertainties on each of these \LambdaCDM parameters \citep{planckcollaboration_2018_VI}. Initially, priors of $10\sigma$ were used, but forecasts were found to be heavily prior driven for $f_b$, $\Omega_m$ and $\sigma_8$, so the priors were loosened until they reached equilibrium values, which occurred at $\sim 100\sigma$.

Priors on the covariance between \LambdaCDM parameters could be drawn from \CMB experiments, however the covariances between \LambdaCDM and \LIM or \ping model parameters are not known, so for the first instance, we assume no direct covariances and consider two prior scenarios, one with loose priors and one with aggressive priors.

We further consider the \ping model parameters: amplification, $\mathcal{Q}$, characteristic wave-number, $k_\mathrm{pulse}$, effective mass term, $m_\lambda$ and the mass of the transverse inflationary field, $m_\chi$. The fiducial values of the \ping model parameters are chosen such that the power spectrum feature is in the range that is observable to \CCAT, $k_\mathrm{pulse}=(\frac{1}{6}~\mathrm{Mpc})^{-1}$, and such that the effect is moderate at this scale, $\log_{10}\mathcal{Q}=-13.625$. Other fiducial values of $\mathcal{Q}$ were investigated and the shape of the relative entropy curve was largely unchanged. No prior is placed on the characteristic wavenumber, $\kpulse$, as it is unconstrained. The prior on $\mathcal{Q}$ is informed by the work in Section \ref{sec:power spectrum} and chosen based on values of $\mathcal{Q}$ that are disfavoured by power spectrum considerations in the range of wavenumbers, $\kpulse$, that the \CCAT survey volume is sensitive to.

The fisher matrix is inverted to get covariance, $\mathbf{Cov} = \mathcal{I}^{-1}(\bar{\bm\lambda})$, or with priors, $\mathbf{Cov} = [\mathcal{I}(\bar{\bm\lambda})+\mathcal{I}_\mathrm{prior}(\bar{\bm\lambda})]^{-1}$, which has diagonal, $\mathrm{diag}(\bm{\sigma})$. These are then normalized to give the correlation matrix, $\mathrm{Corr}_{ij} = \frac{ \mathrm{Cov}_{ij} }{ \sqrt{ \sigma_i \sigma_j} }$, with unit diagonals. Figure \ref{fig:marginalized DSS and S2 forecasts} shows confidence ellipses from Fisher analysis with the \CCAT \DSS noise model as solid blue contours and with the \Stwo \LIM noise model as orange contour lines. The fiducial values of each parameter are shown above each \PDF along the diagonal of the corner plot. The uncertainty displayed is the forecasted $1\sigma$ uncertainty on the parameter values. Parameters that were prior driven are marginalized over, the full corner plot is shown in Appendix \ref{app:full forecasts}. The marginalized $1\sigma$ posterior for each model parameter is also listed in Table \ref{tab:LIM Srel params}.

We observe some cross-correlation between Hubble's constant, $H_0$, and $\sigma_Z$ as well as with the \ping model parameters, which is unsurprising as $H_0$ measures the current rate of expansion of the universe, which modulates the dominance of gravity in the formation of structures. A higher $H_0$ means gravity is less dominant compared to the cosmological constant, $\Lambda$, driving accelerated expansion, and there will be less clustering, and therefore fewer and smaller \DM halos. Since the \LIM signal traces halos, and particularly the most massive halos, therefore a larger metallicity scatter, $\sigma_Z$, would be needed to compensate for a given discrepancy in the \LIM signal intensity. Similarly, stronger \ping effects and thus a higher $\mathcal{Q}$ could compensate for the weaker gravity implied by a high value of $H_0$. There is a strong correlation between $\kpulse$ and $\mathcal{Q}$, which makes sense as both parameters affect the amplitude, $\mathcal{D}_0$, of the \ping power spectrum feature as was demonstrated by Figure \ref{fig:CobayaMCMC}. Given these correlations, it is unsurprising that there would be some correlation between $\kpulse$, $H_0$ and $\sigma_Z$.

The \Stwo \LIM forecast localizes the \cii amplitude $\alphacii$ to within 300\% of the fiducial value, and the metallicity scatter to within 100\%. The \CCAT \DSS forecast is less sensitive by a factor of less than 2. We therefore cannot conclude that sub-millimetre surveys will put robust constraints on the cosmological line emission model, at least at the level of one-point statistics of autocorrelation studies. More constraining power on \LIM models may be found in statistics beyond the one-point \VID, or in cross-correlation studies that can better evade foregrounds. Folding in constraining power obtained from the \HMF, and \ping power spectrum could improve matters further. Encouragingly, the \LIM model parameters are not strongly correlated with \LambdaCDM or \ping model parameters. Thus, once observations of the \LIM signal are made, one can distinguish a situation in which the \LIM model does not accurately fit the data from one that suggests new physics.

Neither the \Stwo \LIM or \CCAT \DSS noise models prove to be particularly sensitive to \LambdaCDM parameters, with most of the parameters that were considered in this analysis being strongly prior-driven. Any constraints on $\Omega_m$, $f_b$ or $\sigma_8$ from these methods would not be competitive with other measures. For $H_0$ on the other hand, the forecast predicts a data-driven confidence ellipse that is competitive with other measures. This constraint does come, albeit, from fitting to a \LambdaCDM model with \CMB-informed priors, so this does not constitute an independent measurement of the Hubble constant. One could run comparative studies of \pkp and \ws simulations using a Hubble constant in the regime predicted by supernova studies, $H_0\sim73 ~ \mathrm{km} ~ \mathrm{s}^{-1} ~ \mathrm{Mpc}^{-1}$ \citep{Riess_2022,riess_2024}, or those from \BAO \citep{guo_2024} and stellar population studies, $H_0\sim70 ~ \mathrm{km} ~ \mathrm{s}^{-1} ~ \mathrm{Mpc}^{-1}$ \citep{Freedman_2021}, and consider relative likelihoods of either model against \LIM observations.

These forecasts place substantial constraints on the characteristic \ping wavenumber, $\kpulse$, with $1\sigma$ posteriors of 1 to 2 orders of magnitude for \Stwo \LIM and \CCAT \DSS models respectively. This is an impressive constraint on the otherwise completely free parameter, but it should be stressed that this Fisher forecast assumes each posterior autocorrelation is described by a Gaussian \PDF. Conversely, we showed in the \MCMC analysis of a \ping-like power spectrum feature that constraints from the power spectrum result in a bimodal $\kpulse$ \PDF. Whether that is also the case for field-level \ping propagated to tomographic mocks of the cosmological \cii signal would require further study, perhaps by training an emulator on \ping mocks and deriving a full set of posteriors from \MCMC. The constraints placed on the \ping amplification factor, $\mathcal{Q}$, would likely also improve with \MCMC-derived posteriors, as again, \MCMC analysis on the \ping-like power spectrum feature suggest autocorrelations of $\mathcal{D}_0$ do not form neat elliptical posteriors, and thus the related $\mathcal{Q}$ would likely behave similarly.

While the remaining \LambdaCDM parameters, $f_b$, $\Omega_m$ and $\sigma_8$, were found to be prior driven in this analysis, this should not be misunderstood to mean that \pkp and \ws, or this relative entropy-based ansatz, are insufficient tools for studying the \LambdaCDM model generally. Using \ws response functions to other observables, or \ws mocks derived from \pkp catalogues at scales and/or epochs not captured by this \CCAT-focussed work, posteriors could be derived for \LambdaCDM parameters. In a recent work for instance, sensitivity to \LambdaCDM parameters is demonstrated in cross-correlations of the \tSZeffect and galaxy catalogues when using oriented stacking on clusters in \ws \tSZ mock catalogues \cite{lokken_2025}. As was the case with \LIM model parameter constraining power, improvements could likely be made by incorporating statistics beyond the one-point, or in cross-correlation studies. Furthermore, we stress that constraints on \ping reported in this work are not all-encompassing. \ping effects in the strong-instability regime, or on \LSS beyond the regime probed by \CCAT remain largely unconstrained and an exciting area for future research.

The \Stwo \LIM forecast indicates that the characteristic scale of \ping, if present, could be constrained at a 1-$\sigma$ level to within a few orders of magnitude. The \CCAT \DSS forecast has less constraining power by a factor of about 2. For a more sophisticated analysis made in conjunction with power spectrum constraints, one might see a marked improvement in constraining power. Consider the posteriors shown in Figure \ref{fig:CobayaMCMC}. The power spectrum strongly rules out \ping with characteristic scale around the \BAO scales, $10^{-2} \lesssim k/(h^{-1}\mathrm{Mpc}) \lesssim 2\times10^{-1}$. Thus, coupled with a \Stwo \LIM survey, a detailed study of \CMB foregrounds and power spectra could prove a powerful constraint on the parameter space of this new inflationary effect.

\section{Conclusions}\label{sec:conclusions}

In this work, we demonstrate that the \pkp and \ws simulations are a suitable tool to produce the many mocks necessary for a thorough investigation of the \LIM, \LambdaCDM and potentially even some \png model spaces. We reinforce the \VID as a powerful summary statistic, demonstrating that it can be used as the basis for forecasts of upcoming \LIM surveys. Using the relative entropy formalism introduced by \citeHCBL, we show that the success of the \VID approach lies in the information-richness of intensity space.

While no direct observation of cosmological \LIM signals have yet been observed, clever stacking methods, such as those done in \cite{dunne_2024,dunne_2025}, and cross-correlations \citep{chung_2019b} will likely soon take these first steps. As has been shown for \ws mocks and \CMB foregrounds, observations of the cosmological \LIM signal could be validated on \ws \LIM mocks. The techniques used in this work are transferable, for instance the \ws simulations already contain a host of response functions for \LIM and \CMB foreground mock maps. The \pkp-\ws simulation pipeline is therefore a strong candidate for extensively use in making mock maps needed in the study of cross-correlation between surveys and across emission lines.

We place new constraints on \ping introduced by \citeMBB, a new and largely unconstrained type of \png. \ping arise generically in multi-field inflation, rather than being specific to a single niche model. Searches for \ping could provide new insights into inflation. We establish the difference between power spectrum features in Gaussian and non-Gaussian fields and motivate the need to look beyond at field-level effects in cases where \png generates features in the power spectrum. We further demonstrate that constraints can be placed on the \ping parameter space using mock halo catalogues from the \pkp simulations, which have support for implementation of full, field-level \ping effects. Relative entropy techniques show the richness of information in halo mass space accessible for these sorts of surveys. We leave comparison between \pkp halos and related observables such as \LRGs to future work.

Finally, we demonstrate the power of \ws in producing forecasts for upcoming surveys. Fisher analysis shows that in the tomographic \cii intensity, effects originating from the \LIM model, cosmology, or potentially even early-universe physics such as \ping (at least for \Stwo \LIM surveys), can be distinguished from one another. The first major implication of this finding is that it will immediately be clear whether any observed deviations from the expected cosmological \cii signal are a result of new fundamental cosmology or inaccuracies in the simple galaxy-halo connection ansatz that we employ in this work. While the one-point statistics of a single \LIM observation prove insensitive to \LIM model parameters, $\mathrm{LIM}\times\mathrm{LIM}$, $\mathrm{LIM}\times P(k)$ and $\mathrm{LIM}\times\mathrm{LSS}$ cross-correlations and higher-order statistics will likely prove more fruitful given the constraining power we demonstrate in \HMF and power spectra. Therefore, as soon as observations are made, we can test and improve our understanding of the the astrophysics that produces the cosmological \LIM signal and update our response functions accordingly. This distinguishability further implies that we will be able to quickly start doing cosmology with \LIM survey data once new telescopes such as \CCAT become operational.

The other major implication of this finding is that \Stwo \LIM surveys will be sensitive to \ping, opening the door to the study of an entirely new parameter space of multi-field inflation models. Sensitivities will only improve when we consider cross-correlations with other telescopes and lines, breaking degeneracies we observed between certain model parameters. This strongly motivates the implementation of further \ws \LIM mocks featuring \ping.

The \ping model used in this work is by no means all-encompassing. The \ping studied in this work for instance are tuned to have characteristic scales in the regimes that \CCAT \cii surveys will be sensitive to, and the approximate methods used in this work to model the strong instability regime leave out much of the \ping parameter space. We have only just begun to explore the effects of \ping, and the parameter space remains wide open. Given these results, we propose \ping as a new science case for upcoming \LIM surveys and cross-correlation efforts.

\acknowledgments

Research in Canada is supported by NSERC and CIFAR. \pkp calculations were performed on the Niagara supercomputer at the SciNet HPC Consortium. SciNet is funded by the Canada Foundation for Innovation under the auspices of the Digital Research Alliance of Canada, the Government of Ontario, Ontario Research Fund – Research Excellence, and the University of Toronto. \cii \ws mocks and other parts of these calculations were performed on high-performance computing nodes at CITA; the authors thank John Dubinski, MJ Huang, Jay Li and Daniel Xiao for tireless maintenance of these resources.

This research was supported in part by grant NSF PHY-2309135 to the Kavli Institute for Theoretical Physics (KITP).

NJC thanks Maria Mylova and Adam Smith for discussions of the \png and cosmological \MCMC codes, and Azadeh Moradinezhad Dizgah for sharing insights into \png and power spectrum features.

PH acknowledges support from a doctoral scholarship from NSERC.

This research made use of Astropy,\footnote{\href{https://www.astropy.org}{https://www.astropy.org}} a community-developed core Python package for astronomy \citep{astropy_2013,astropy_2018,astropy_2022}, as well as the NumPy \citep{harris_2020} and SciPy \citep{virtanen_2020} packages.

\appendix

\section{The \ping potential}\label{app: ping potential}

In the lattice simulations by \citeMBB, a generic multi-field inflation template is considered with an inflaton-like field, $\phi$, parallel to the mean flow of the effective potential of inflation. The inflaton, $\phi$, is assumed to be slowly rolling, and a second transverse field, $\chi$, is sitting at the minima of its potential, which is perpendicular to the mean flow. The potential has an feature giving rise to intermittent symmetry breaking in the $\phi$ and $\chi$ trajectories. Outside the instability, the potential is taken to be quadratic, $V_0(\phi,\chi) = \frac{1}{2} m_\phi^2 \phi^2+\frac{1}{2} m_\chi^2 \chi^2$, where $m_\phi$ and $m_\chi$ are the respective free-field masses of the inflationary scalar fields. 
The quadratic potential is chosen as it has transverse symmetry, which will be broken during when trajectories pass through the potential feature. This should be understood as a parameterization of the potential in the neighbourhood of the instability only, not a description of the potential for the entirety of inflation. Therefore, measurements of the spectral scalar index $n_s$ and tensor-to-scalar ratio $r$ that rule out a quadratic inflationary potential \cite{Paoletti_2022} do not apply. We model the symmetry breaking as a small quartic perturbation to the quadratic potential,
\begin{equation}
    \Delta V(\phi,\chi)=
    \begin{cases}
        \frac{\lambda_\chi}{4}
        \left[\left(\frac{\phi-\phi_p}{\phi_w}\right)^2-1\right]^2
        \left[\left(\chi^2 - v^2\right)^2 - v^4\right], & \forall |\phi-\phi_p| < \phi_w,\\[6pt]
        0, & \forall |\phi-\phi_p| \ge \phi_w.
    \end{cases}
\end{equation}
\citeMBB note that the precise mathematical form of the potential feature has little qualitative impact on the resulting \nong so long as it is intermittent and gives rise to a negative effective mass of the transverse field, $\chi$. $\Delta V$ is quartic in $\chi$ with a local maximum, $\Delta V(\phi,0)=0$, and a pair of global minima at points $\chi=\pm v$, the vacuum expectation value of $\chi$ during the instability. The depth of the minima is extremized at $\phi=\phi_p$. 

In the strong instability limit, $\chi$ trajectories do not reach $\pm v$ before the end of the instability, thus $|\chi| \ll v$. This limit is valid if $v \ll \phi_w$, and the non-zero potential perturbation becomes,
\begin{align}\label{eq:strong instability potential response}
    \Delta V (\phi,\chi)
    &\simeq -\frac{1}{2} m_\lambda^2 \chi^2 \left[ \left( \frac{\phi-\phi_p}{\phi_w} \right)^2 -1 \right]^2,
    \qquad\forall |\phi-\phi_p| < \phi_w,
\end{align}
where $m_\lambda \equiv \sqrt{\lambda_\chi} v$. When the instability is maximal, $\Delta V(\phi_p,\chi) = -m_\lambda^2 \chi^2/2$, which plainly contributes a negative effective squared mass to the $\chi$ field, $m_{\chi,\mathrm{eff}}^2 \equiv m_\chi^2-m_\lambda^2$. The Klein-Gordon equation, $\square\psi+\partial_\psi V=0$ \cite{Klein_1926,Gordon_1926}, has Fourier transform, $\ddot{\tilde{\chi}} + (k^2 + m_{\chi,\mathrm{eff}}^2)\tilde{\chi}=0$, at $\phi=\phi_p$. Solving, we see that if the effective mass of the transverse inflationary field $m_{\chi,\mathrm{eff}}^2<0$, then modes of $\tilde{\chi}$ with $k^2 < - m_{\chi,\mathrm{eff}}^2$ grow exponentially with time while $k^2 > - m_{\chi,\mathrm{eff}}^2$ preserves the harmonic oscillator equation, so those modes continue to fluctuate. Thus, an instability with $m_\chi^2 < 0$ requires $m_\chi < m_\lambda$. 

Furthermore, $k<k_{\mathrm{inst},p}$ is said to be the instability band at $\phi=\phi_p$ with $k_{\mathrm{inst},p} \equiv ( m_\lambda^2 - m_\chi^2)^{1/2}$. From equation (\ref{eq:strong instability potential response}), we can see that the effective mass deviates most from $m_\chi$ at $\phi_p$ so only modes with $k<k_{\mathrm{inst},p}$ will grow during the instability, but the largest modes near $k_{\mathrm{inst},p}$ only grow while the inflaton, $\phi$, is in the vicinity of $\phi_p$, where the instability in the potential is at its most dramatic.

In this work, the strong instability limit is used with $v = 10^{-6} M_{Pl}$. The duration $\phi_w$ is chosen such that a few $e$-foldings occur during the instability. This ensures that an insufficient amount of energy is transferred from the potential to kinetic and gradient terms for the symmetry breaking to bring about the end of inflation. The parameter space of \ping is varied and rich in information, in this work we confine ourselves to a single case, but it should not be confused for the only form \ping can take.

\section{Full Fisher Forecasts}\label{app:full forecasts}

\begin{figure}
    \centering
    \vspace{-1em}
    \includegraphics[width=1.0\linewidth]{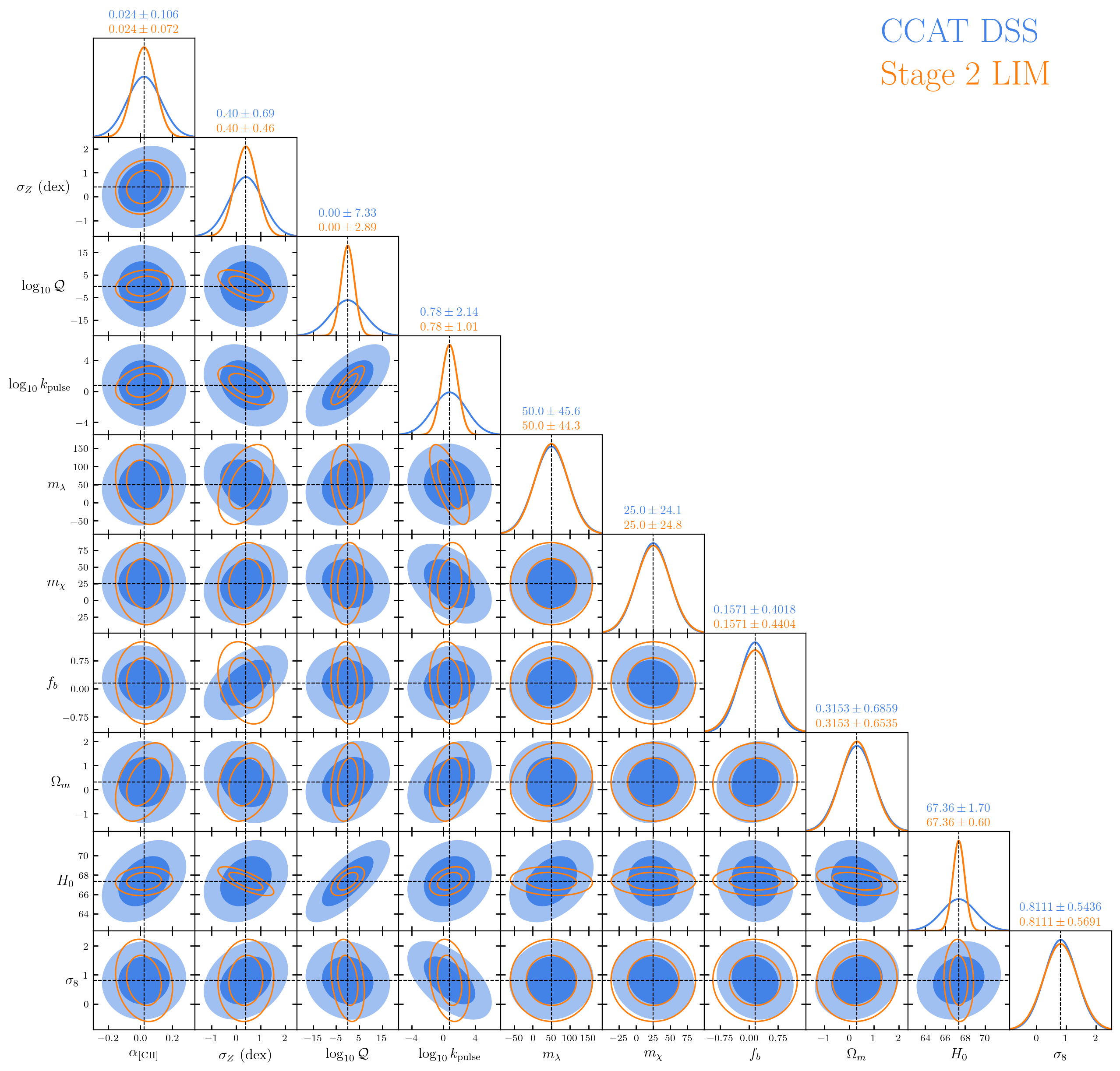}
    \vspace{-0.5cm}
    \caption{$1\sigma$ and $2\sigma$ confidence ellipses with \CCAT \DSS (blue solid ellipses) and \Stwo (orange contour lines) noise models. All parameters varied in the Fisher analysis are shown, including those with prior-driven forecasted \PDF. The mean and $1\sigma$ uncertainties in each \PDF are shown above the diagonal panels.}
    \label{fig:DSS and S2 forecasts}
\end{figure}

In this appendix, we show the full \cii \LIM Fisher forecasts for \CCAT \DSS and \Stwo \LIM surveys without marginalizing over any parameters as was done in Figure \ref{fig:marginalized DSS and S2 forecasts}. Priors on the \ping model masses, $m_\lambda$ and $m_\chi$, are informed by the early-universe work \cite{morrison2024a}. These parameters are discussed in more detail in Appendix \ref{app: ping potential}, they do not have as strong an effect on the \ping feature amplitude as the amplification factor, $\mathcal{Q}$, and it is thus not surprising that high-precision constraints on masses are not achieved. Because we are more concerned with \ping-like scenarios than a specific model of the inflationary potential, this lack of constraining power is of little significance.

Using a 10-$\sigma$ prior for $H_0$ was sufficient to avoid a prior-driven forecast. However, assuming a 10-$\sigma$ prior on each of the other \LambdaCDM parameters, $f_b$, $\Omega_m$ and $\sigma_8$, resulted in strongly prior-driven forecasted probabilities. As shown in Table \ref{tab:LIM Srel params}, a much looser set of priors are used in Figure \ref{fig:DSS and S2 forecasts}, still resulting in largely prior-driven probabilities, with $1-\sigma^2_\mathrm{forecast}\sigma^{-2}_\mathrm{prior} \lesssim 0.2$. These constraints are orders of magnitude weaker than \CMB constraints, indicating that \DSS and \Stwo surveys of the cosmological \cii signal will likely not be sensitive to these parameters. We therefore marginalize over them in our analysis in Section \ref{subsec:fisher}.


\bibliographystyle{JHEP} 
\bibliography{bib} 

\end{document}